\documentclass[12pt]{article}
\pdfoutput=1
\usepackage{amsfonts}
\usepackage{mathptmx}
\usepackage{amsmath}
\usepackage[top=2cm,left=1cm,right=1cm,bottom=2cm,includehead,includefoot]{geometry}
\usepackage{cite}
\usepackage{color}
\usepackage{setspace}
\usepackage[T1]{fontenc}
\usepackage[latin1]{inputenc}
\usepackage{verbatim}
\usepackage{subfigure}
\usepackage{float}
\usepackage[pdftex]{graphicx}
\usepackage{graphicx}

\begin{document}

\title{Observational constraints of the Gravitational Waves in the Brans-Dicke Theory}
\author{R. C. Freitas\thanks{e-mail: \texttt{rc\_freitas@terra.com.br }} \ and S. V. B. Gon\c{c}alves\thanks{e-mail: \texttt{sergio.vitorino@pq.cnpq.br}} \\
\mbox{\small Universidade Federal do Esp\'{\i}rito Santo, Centro de Ci\^{e}ncias Exatas, Departamento de F\'{\i}sica}\\
\mbox{\small Av. Fernando Ferrari, 514 - Campus de Goiabeiras, CEP
29075-910, Vit\'oria, Esp\'{\i}rito Santo, Brazil}}
\date{\today}
\maketitle

\begin{abstract}
         We investigate the quantum origin of the primordial cosmological gravitational waves in the Brans-Dicke theory. We compute the number
of gravitons $N_k$ produced during inflation and the observables: power spectrum $P_T$, spectral index $n_T$ and energy density 
$\Omega_k$. By comparison with General Relativity we see that the results for both theories are the same for the case of the particles number
$N_k$ and for the case of the power spectrum $P_T$ and the energy density $\Omega$ only when $\omega<10$. For the spectral index $n_T$ we found that when the Brans-Dicke coupling parameter $\omega$ is bigger than unity the spectral index approximates the expression found in the General Relativity. This may awake us to the possibility that $\omega$ varies with the cosmological scale.

\vspace{0.7cm}

\par
KEYWORDS: scalar-tensor gravity, gravitational waves
\par
\vspace{0.7cm}
\par
PACS numbers: 04.30.-w, 98.80.-K

\end{abstract}

\section{Introduction}
\label{intro}
\par
The gravitational waves are tensorial fluctuations in the metric of spacetime. This particular perturbation is not explicitly coupled with the energy density and pressure of the matter of the Universe and does not contribute to the gravitational instability that generates the cosmological structures we see today. On the other side his study is of great interest because it supplies the specific signature of the metric theory of gravity. Nevertheless, they must have left special signatures in the polarization of the CMB anisotropies \cite{paul1,paul2, mass}. Moreover, since the gravitational waves were generated in the early Universe, before the time of last scattering of CMB photons, they can be a window to the primordial phase of evolution of the Universe. These waves are predicted by Einstein's theory of General Relativity, but they still have to be directly detected. Great efforts are been done in this sense and there is hope that a new generation of experiments in space may allow this detection in some years \cite{gwobs}. The gravitational waves spectrum frequency extends over a wide range of interest, from $10^{-18}Hz$ to $10^{8}Hz$\footnote{The wave number interval is from $10^{-26}m^{-1}$ to $10^{-16}m^{-1}$.}, depending of the sources that generate those waves. Many works have been made in order to identify specific signatures of cosmological models in the spectra of gravitational waves, for example, in the case of quintessence model \cite{uzan} and string cosmology \cite{ungarelli,sanchez,gasperini}.
\par
The existence of a classical scalar field in the nature has been considered in many theories of gravitation that present alternatives to General Relativity. The prototype of scalar-tensor theories is the Brans-Dicke theory \cite {1, 2, 3, 4}, whose Lagrangian \textcolor{red}{in the Jordan frame} is given by
\begin{equation}
\label{lagr}
\mathcal{S} = \frac{1}{16\pi}\int d^{4}x\sqrt{-g}\biggl[\phi R - \omega\biggl(\frac{\phi_{,\mu}\phi_{,}^{\mu}}{\phi}\biggr) + 16\pi\mathcal{L}_{mat}\biggr]\quad,
\end{equation}
where $\phi$ is a scalar field that couples to gravity through the parameter $\omega$, called Brans-Dicke parameter and $\mathcal{L}_{mat}$ is the matter Lagrangian.
\par
The Brans-Dicke theory, as others scalar-tensor theories, can be formulated in the Einstein or in the Jordan frames, which are conformally related. Which conformal frame is the physical one is a very contentious issue \cite{faraoni}. The most common argument against the Jordan frame found in the literature is that the scalar field energy density can assume negative values. This happens because of the nonmininal coupling between the scalar field and geometry. The terms with the second covariant derivatives of the scalar field in the Brans-Dicke field equations contain the connection $\Gamma^{\alpha}_{\mu\nu}$ and therefore part of the dynamical description of gravity. But the energy density can be made nonnegative with the help of a new connection $\tilde{\Gamma}^{\alpha}_{\mu\nu}$ \cite{santiago} which is also associated with the physical metric $g_{\mu\nu}$.  Based on the considerations presented in the previous reference  we work here considering Jordan frame as the physical frame.
\par
One can expect that the presence of the scalar field leads to different predictions about the results obtained in General Relativity. The standard cosmological scenario given by General Relativity shows some spectacular successes but also have some puzzles, such as the flatness problem, the horizon problem and structure formation problems \cite{5}.
\par
 The value of the parameter $\omega$ can be limited by local physics experiments like Cassini experiment \cite{cassini, will}: it should be huge today, the order of $\omega = 40000$ to satisfy the experimental tests of General Relativity. This result reduced the interest in the Brans-Dicke theory because when $\omega\rightarrow\infty$ in the motion equations we obtain the General Relativity. The situation has changed with the proposal of the extended inflation \cite{6, 7}. The idea is to take advantage of the time dependence of gravitational constant $G$ to solve the bubble nucleation problem that arises in tradicional inflation. In the de Sitter phase, Brans-Dicke theory predicts power-law inflation instead of exponential; in fact, following \cite{6}, at the beginning of inflation the Brans-Dicke solutions (for $p = -\rho$) approaches the Einstein-de Sitter solution. In the second stage of the inflation, both the scalar field and the scale factor grow by power law rather than exponential. This feature prevents the so-called graceful exit problem. However, in order to work, the parameter must be $\omega\approx 24$, contradicting local observations. This drawback can be overcome through a generalization of the original Brans-Dicke theory, allowing the parameter $\omega$ to be a function of the field $\phi$ itself.
\par
In 1999, the SN Ia observations \cite{riess} showed that the universe is currently undergoing accelerated expansion. A possible theoretical explanation for this acceleration is the vaccum energy with negative pressure, called dark energy (that violates the strong energy condition). There are many dark energy models, for example quintessence \cite{quinte}, k-essence \cite{k}, phantom \cite{fantasma} and Chaplygin gas \cite{chap}. The simplest candidate for the dark energy is the cosmological constant $\Lambda$ whose equation of state is $p = -\rho$. But there are some problems with this alternative: why the cosmological constant is so small, nonzero, and comparable to the critical density at the present? So, it is interesting to study dynamical dark energy models like the Brans-Dicke theory of gravity as a possible theory of k-essence coupled to gravity since it involves the simplest form of non-linear kinetic term for the Brans-Dicke scalar field $\phi(t)$ \cite{rose}.
\par
This revival of the Brans-Dicke theory leads us to ask if the generation of gravity waves can be modified through the introduction of this scalar field. Here, we propose to study the evolution of tensorial fluctuation in the traditional Brans-Dicke theory and its quantum origin and calculate the observables that can be compared with the observational data. The main interest in the presence of the scalar field is the possibility that it can change the evolution of gravity waves in comparison with the results obtained with General Relativity \cite{barrow}.
\par
This paper is organized as follows. In section II we show the field equations, the equations of motion of the Brans-Dicke theory and the background solutions. In section III we obtain the perturbed equations of gravity waves and their solutions in terms of Hankel functions in the primordial phase of the universe. In section IV the quantum formulation of the gravitational waves in the Brans-Dicke theory is obtained. In section V we calculate the observables of the Brans-Dicke theory. Finally, in section VI we discuss the results and show our conclusions.
\par
We use in this article standard tensor notation: the greek indices run from zero to three; the latin indices run from one to three; the space-time metric is taken to have signature $+2$, or $(-,+,+,+)$; covariant derivatives are indicated by semicolons; partial derivatives are indicated by commas and the scalar field $\phi$ is a time function.

\section{Background solutions}
\label{section_model}

We assume the background Universe is spatially flat, homogeneous and isotropic, {\it i.e.}, it is described by the Robertson-Walker metric
\begin{equation}
               \label{metric}
               \mathrm{d}s^2 = -\mathrm{d}t^2 + a^{2}(t)\left(\mathrm{d}x^2+\mathrm{d}y^2+\mathrm{d}z^2 \right)
               \hspace{0.2cm} \textrm{,}
\end{equation}
where the $a(t$) is the scale factor of the Universe, and $c = 1$. The energy-momentum tensor of the background matter takes a perfect fluid form
\begin{equation}
               \label{eq:fluidoperfeito}
               T^{\mu\nu}=\left( \rho + p\right)u^\mu u^\nu +p g^{\mu\nu}
               \hspace{0.2cm} \textrm{,}
\end{equation}
with an equation of state $p = \alpha\rho$. In these expressions $\rho$ is the matter density, $p$ is the pressure and $\alpha$ is a constant.
From the action (\ref{lagr}) we obtain the field equations
\begin{eqnarray}
\label{field1}
R_{\mu\nu} - \frac{1}{2} g_{\mu\nu} R &=& \frac{8\pi}{\phi} T_{\mu\nu} + \frac{\omega}{\phi^2}\biggl( \phi_{;\mu}\phi_{;\nu} - \frac{1}{2} g_{\mu\nu}\phi_{;\rho}\phi_{;}^{\rho}\biggr) + \frac{1}{\phi}\biggl(\phi_{;\mu ;\nu} - g_{\mu\nu}\Box\phi\biggr)\quad,\\
\label{field2}
\Box\phi &=& \frac{8\pi}{3 + 2\omega} T\quad.
\end{eqnarray}
\par
Inserting the metric (\ref{metric}), into equations (\ref{field1}) and (\ref{field2}) we obtain the equations of motion
\begin{eqnarray}
\label{motion1}
- 3\frac{\ddot a}{a} &=& \frac{8\pi}{\phi}\rho\biggl(\frac{2 + \omega + 3\alpha + 3\alpha\omega}{3 + 2\omega}\biggr) + \omega\frac{\dot\phi^2}{\phi^2} + \frac{\ddot\phi}{\phi}\quad, \\
\label{motion2}
\ddot\phi + 3\frac{\dot a}{a}\dot\phi &=& \frac{8\pi}{3 + 2\omega}\rho(1 - 3\alpha)\quad.
\end{eqnarray}
\par
The above equations must be supplemented by a conservation equation of the energy-momentum tensor
\begin{equation}
T^{\mu\nu}_{;\nu} = 0\quad.
\end{equation}
\par
When $\mu = 0$ we obtain the energy conservation
\begin{equation}
\label{energia}
\dot\rho + 3\frac{\dot a}{a}\rho(1 + \alpha) = 0\quad,
\end{equation}
while, for $\mu = j$ we get the condition
\begin{equation}
\label{cm}
\rho_{,i} = 0\quad.
\end{equation}
\par
Background solutions \cite{plinio} can be obtained assuming that the scale factor $a(t)$ and the scalar field $\phi(t)$ have a power-law form
\begin{equation}
a(t)\propto t^r~~~~ ,~~~~\phi(t)\propto t^s\quad.
\end{equation}
By direct substitution of the above relation in the equations (\ref{motion1}), (\ref{motion2}), (\ref{energia}) and (\ref{cm}), we have
\begin{eqnarray}
\label{soba1}
- 3r(r - 1) &=& 8\pi\rho t^{2-s}\biggl(\frac{2 + \omega(1 + 3\alpha) + 3\alpha}{3 + 2\omega}\biggr) + s^2 (1 + \omega) - s\quad,\\
\label{soba2}
s^2 + s(3r - 1) &=& \frac{8\pi}{3 + 2\omega}\rho t^{2-s}(1 - 3\alpha)\quad,\\
\label{soba3}
\dot\rho + ~3\frac{r}{t}\rho(1 + \alpha) &=& 0\quad,\\
\label{soba4}
\rho_{,i} &=& 0\quad.
\end{eqnarray}
\par
In what follows we shall consider the inflation case ($\alpha = -1$), the radiation phase ($\alpha = 1/3$) and the dust era ($\alpha = 0$). In the first scenario we have a drastic expansion of the Universe during the early
period of the Big Bang. In this case, the background solutions are
\begin{equation}
\label{infla}
s = 2~~, ~~~~ r = \omega + \frac{1}{2}\quad.
\end{equation}
In the radiation phase, with $\alpha = 1/3$, we have the same background solution that the General Relativity
\begin{equation}
\label{rad}
s = 0~~, ~~~~ r = \frac{1}{2}\quad.
\end{equation}
In the dust era ($\alpha = 0$) we obtain
\begin{equation}
\label{poe}
s = 2 - 3r~~, ~~~~ r = 2\frac{1 + \omega}{4 + 3\omega}\quad,
\end{equation}

\section{The gravitational waves equation in the Brans-Dicke theory}
\label{section_h}

Cosmological gravitational waves are obtained by means of a small correction $h_{ij}$ in equation (\ref{metric}), which represents the metric. Hence, the general expression of the metric (\ref{metric}), related to the unperturbed metric, is replaced by
\begin{equation}
               \label{metric1}
               \mathrm{d}s^2 = -\mathrm{d}t^2 + \left[a^{2}(t)\delta_{ij}+h_{ij}\right]\mathrm{d}x^i \mathrm{d}x^j
               \hspace{0.2cm} \textrm{.}
\end{equation}
Appling the metric (\ref{metric1}) in the field equation (\ref{field1}) the resulting expression is
\begin{equation}
\label{gw1}
               \ddot{h}_{ij}+\left(\frac{\dot{\phi}}{\phi} -\frac{\dot{a}}{a}\right)\dot{h}_{ij}+
               \left[\frac{k^2}{a^2}-2\frac{\ddot{a}}{a} -2\frac{\dot{a}\dot{\phi}}{a\phi} \right]h_{ij}=0
               \hspace{0.2cm} \textrm{,}
\end{equation}
where $k$ is the wavenumber, the dots indicate cosmic time derivatives and we have written $h_{ij}(t,\vec x) =  h(t)Q_{ij}$, where $Q_{ij}$ are the eigenmodes of the Laplacian operator, such that $Q_{ii} = Q_{ki,k} = 0$.
\par
Performing the transformation from the cosmic time to the conformal time, $a(t)~d\eta = dt$, and representing the derivatives with respect to $\eta$ by primes, equation (\ref{gw1}) assumes the form
\begin{equation}
\label{gw2}
h_{ij}''+\left(\frac{\phi'}{\phi}-2\frac{a'}{a} \right)h_{ij}'+  \left[k^2-2\frac{a''}{a}+2\frac{a'^2}{a^2}-2\frac{a'\phi'}{a\phi} \right]h_{ij}=0
\hspace{0.3cm} \textrm{.}
\end{equation}
\par
To resolve this differential equation we need to perform the Fourier transformation
\begin{equation}
               \label{eq:perturbfourier}
h_{ij}(\vec{x},\eta)=\sqrt{16\pi}\sum_{\lambda=\otimes,\oplus}\int{\frac{\mathrm{d}^{3}k}{(2\pi)^{3/2}}
               \epsilon_{ij}^{(\lambda)}(\hat{k})\frac{a(\eta)\mu_{(\lambda)}(\eta)}{\sqrt{\phi(\eta)}}e^{-\dot{\imath}\vec{k}\cdot\vec{x}} }
               \hspace{0.3cm} \textrm{,}
\end{equation}
where the polarization tensor $\epsilon_{ij}^{(\lambda)}(\hat{k})$ can be decomposed as $\epsilon_{ij}^{(\lambda)}(\hat{k})\epsilon^{ij(\lambda')}(\hat{p})=2\delta^{\lambda \lambda'}\delta^{(3)}(\vec{k}-\vec{p})$
to the two polarization states $\otimes$ e $\oplus$. Then, we have
\begin{equation}
               \label{eq:mudif1}
\mu_{(\lambda)}''(\eta) + \biggl[k^2 + \frac{a''}{a}+\frac{1}{2}\frac{\phi''}{\phi} - \frac{1}{4}\left(\frac{\phi'}{\phi}\right)^{2}+\frac{a'\phi'}{a\phi}\biggl]\mu_{(\lambda)}(\eta)=0
               \hspace{0.2cm} \textrm{.}
\end{equation}
In this way, we rewrote the equation (\ref{gw2}) in terms of a simple harmonic oscillator.
\par
We can express the equation (\ref{eq:mudif1}) in terms of the new parameter $\Phi(\eta)\equiv a(\eta)\sqrt{\phi(\eta)}$. Hence, the equation (\ref{eq:mudif1}) becomes
            \begin{equation}
               \label{eq:mudif}
               \mu_{(\lambda)}''(\eta)+\left[k^2 -\frac{\Phi''(\eta)}{\Phi(\eta)} \right]\mu_{(\lambda)}(\eta)=0 
               \hspace{0.3cm} \textrm{.}
            \end{equation}   
         \par
With the background solutions of the scale factor and the scalar field we get
$$
\label{eq:potencial}
\frac{\Phi''(\eta)}{\Phi(\eta)}=
\left\{ \begin{array}{lclll}
\frac{F(\omega)}{|\eta|^2}~~\mbox{,} & \mbox{ if } & \alpha = -1   & \mbox{with} &  -\infty  < \eta \leq -\eta_{1}\quad,\\
& & & & \\
0~~\mbox{,} & \mbox{ if } & \alpha = 1/3    & \mbox{with} & -\eta_{1}\leq  \eta \leq \eta_{2}\quad,\\
& & & & \\
\frac{G(\omega)}{(\eta - \tilde{\eta}_{dust})^2}~~\mbox{,} & \mbox{ if } & \alpha = 0  & \mbox{with} &  \eta \geq \eta_{2}\quad,
\end{array}
\right.
$$         
where 
\begin{eqnarray}
              && F(\omega)=\frac{2(3+2\omega)(1+2\omega)}{(1-2\omega)^2}\nonumber\quad, \\
              && G(\omega)=\frac{(3+2\omega)(1+\omega)}{(2+\omega)^2}
               \hspace{0.2cm} \textrm{,}
\end{eqnarray}
and
\begin{equation}
                \tilde{\eta}_{dust}= \eta_{2}-\frac{2+2\omega}{2+\omega}\left[1+\left(\frac{\omega+1/2}{\omega-1/2}\right)
                \left(1+\frac{\eta_{2}}{\eta_{1}}\right)\right]
                \hspace{0.2cm} \textrm{,}
\end{equation}
\par
The solutions, for each stage of evolution of the Universe are
\begin{enumerate}
\item Inflation era $(\alpha = -1)$
             \begin{equation}
                \label{eq:solinflacao}
                \mu(k\eta)=\sqrt{\eta}\left(A_{1}\mathcal{H}^{(1)}_{\nu}(k|\eta|)+A_{2}\mathcal{H}^{(2)}_{\nu}(k|\eta|) \right)
                \hspace{0.1cm} \textrm{,}
             \end{equation}
where $A_{1}$ and $A_{2}$ are integration constants, $\mathcal{H}^{(1)}$ and $\mathcal{H}^{(2)}$ are the Hankel functions of first and second kind, respectively, and $\nu=\sqrt{F(\omega)+1/4}$ is the order of the Hankel functions. 
\item Radiation era $(\alpha = 1/3)$
             \begin{equation}
                \label{eq:solradiacao}
                \mu(k\eta)=B_{1}e^{\dot{\imath}k\eta}+B_{2}e^{-\dot{\imath}k\eta}
                \hspace{0.3cm} \textrm{,}
             \end{equation}
where $B_{1}$ and $B_{2}$ are yet integration constants.
\item Dust era $(\alpha = 0)$
\begin{equation}
\mu(k\eta) = \sqrt{\eta}\biggl[C_{1}\mathcal{H}^{(1)}_{\alpha}\left(k(\eta - \tilde{\eta}_{dust})\right) + C_{2}\mathcal{H}^{(2)}_{\alpha}\left(k(\eta - \tilde{\eta}_{dust})\right) \biggr]
                \hspace{0.2cm} \textrm{,}
\end{equation}
where $C_{1}$ and $C_{2}$ are integration constants and $\alpha=\sqrt{G(\omega)+1/4}$ is the order of the Hankel functions.
\end{enumerate}

\section{The quantum gravitational waves equation in the Brans-Dicke theory}
\label{subsection_BD}

In this process we start with the Lagrangian density of the gravitational waves in the Brans-Dicke theory, given by
\begin{equation}
\label{lagran01}
                  \mathcal{L}=\frac{\phi}{16\pi}\bar{g}^{\mu\nu}\partial_{\mu}h_{ij}\partial_{\nu}h^{ij}
                  \hspace{0.2cm} \textrm{,}
\end{equation}
that leads to two corresponding Hamiltonians
\begin{eqnarray}
                 \label{eq:hamilt1}
                 H^{(1)}&=&-\frac{1}{2}\int{\mathrm{d}^{3}x\left[\tilde{\pi}^2 -2\frac{\Phi'}{\Phi}\mu\tilde{\pi}+
                 \delta^{ij}\mu_{,i}\mu_{,j}\right]} \quad, \\
                 \label{eq:hamilt2}
                 H^{(2)}&=&-\frac{1}{2}\int{\mathrm{d}^{3}x\left[\pi^2 -\frac{\Phi''}{\Phi}\mu^2 +
                 \delta^{ij}\mu_{,i}\mu_{,j}\right]} 
                 \hspace{0.2cm} \textrm{,}
\end{eqnarray}
where $\pi$ and $\tilde{\pi}$ are the canonical momentum. The above Hamiltonians differ each other by addiction or subtraction of a total conformal time derivative.
\par
The expression of the quantum Hamiltonian obtained by the equation (\ref{eq:hamilt2}) for the gravitational waves in the Brans-Dicke theory is constructed by using the canonical commutation relations
\begin{equation}
                 \label{eq:comutacao}
                 \left[\hat{\mu}(\vec{x},\eta),\hat{\pi}(\vec{y},\eta) \right]=\dot{\imath}\delta^{(3)}(\vec{x}-\vec{y})
                 \hspace{0.2cm} \textrm{.}
\end{equation}
\par
So, by transforming the normal modes of oscillation of the gravitational waves in field and momentum operators in the Heisenberg description of quantum mechanics we obtain the Hamiltonian
\begin{equation}
                 \label{eq:hamiltoniana2}
                 H^{(2)}\rightarrow \hat{H}^{(2)}=-\frac{1}{2}\int{\mathrm{d}^{3}x\left[\hat{\pi}^2 
                 -\frac{\Phi''}{\Phi}\hat{\mu}^2 +\delta^{ij}\hat{\mu}_{,i}\hat{\mu}_{,j}\right]} 
                 \hspace{0.2cm} \textrm{,}
              \end{equation}
where the field operator is constructed from its classical field as follows
\begin{equation}
\label{oper01}
                 h_{ij}\rightarrow\hat{h}_{ij}(\vec{x},\eta) = \frac{\sqrt{16\pi}}{a(\eta)\sqrt{\phi(\eta)}}\sum_{\lambda=\otimes,\oplus}\frac{1}{2}
                    \int \frac{\epsilon_{ij}^{(\lambda)}(\vec{k})\mathrm{d}^3k}{(2\pi)^{3/2}}
                    \left(\hat{\mu}_{\vec{k},(\lambda)}
                    e^{-\dot{\imath}\vec{k}\cdot\vec{x}}
                    +\hat{\mu}_{\vec{k},(\lambda)}^{\dagger}e^{\dot{\imath}\vec{k}\cdot\vec{x}}\right)
                    \hspace{0.2cm} \textrm{,}
\end{equation}
with the operator $\hat{\mu}_{\vec{k}}$ given by
              \begin{equation}
                 \hat{\mu}_{\vec{k}}=\hat{a}_{\vec{k}}(\eta_0)f_{k}(\eta)+\hat{a}_{-\vec{k}}^{\dagger}(\eta_0)f_{k}^{*}(\eta)
                 \hspace{0.2cm} \textrm{,}
              \end{equation}
and the function $f_{k}$ satisfying the following equation 
              \begin{equation}
                 f_{k}''+\left(k^{2}-\frac{\Phi''}{\Phi}\right)f_{k}=0 \hspace{0.2cm} \textrm{,}
              \end{equation}
that is the same differential equation that describe the behavior of the classical function $\mu(\eta)$.
\par
Now, to obtain coherent states of the relic gravitons in the Brans-Dicke theory we consider the creation and annihilation operators
\begin{equation}
                 \hat{a}_{\vec{k}}=\sqrt{\frac{k}{2}}\left(\hat{\mu}_{\vec{k}}+\frac{\dot{\imath}}{k}\hat{\pi}_{\vec{k}}\right)
                 \hspace{0.5cm} \textrm{,} \hspace{0.5cm}
                 \hat{a}_{-\vec{k}}=\sqrt{\frac{k}{2}}\left(\hat{\mu}_{\vec{k}}-\frac{\dot{\imath}}{k}\hat{\pi}_{\vec{k}}\right)
                 \hspace{0.2cm} \textrm{,}
              \end{equation}
whose temporal evolution is given by the Heisenberg equation and by the Hamiltonian (\ref{eq:hamilt1}), writen now as operator
\begin{equation}
\label{H01}
                 H^{(1)}\rightarrow\hat{H}^{(1)}=-\frac{1}{2}\int{\mathrm{d}^{3}x\left[\hat{\pi}^2 -\frac{\Phi'}{\Phi}\left(\hat{\mu}\hat{\pi}
                 +\hat{\pi}\hat{\mu}\right)+\delta^{ij}\hat{\mu}_{,i}\hat{\mu}_{,j}\right]} \hspace{0.2cm} \textrm{,}
\end{equation}
with the general solutions given by
\begin{eqnarray}
                 \label{eq:solgeralcoefbogo}
                 \hat{a}_{\vec{k}}(\eta)&=&u_{k}(\eta)\hat{a}_{\vec{k}}(\eta_{0})
                 +v_{k}(\eta)\hat{a}_{-\vec{k}}^{\dagger}(\eta_{0}) \hspace{0.2cm} \quad,\nonumber \\
                 \hat{a}_{-\vec{k}}^{\dagger}(\eta)&=&v_{k}^{*}(\eta)\hat{a}_{\vec{k}}(\eta_{0})
                 +u_{k}^{*}(\eta)\hat{a}_{-\vec{k}}^{\dagger}(\eta_{0}) \hspace{0.5cm} \textrm{,}
\end{eqnarray}
where $\eta_{0}$ is a fixed initial time and the functions $u_{k}(\eta)$ and $v_{k}(\eta)$ behave as
              \begin{equation}
                 \label{eq:transfbogo}
                 \frac{\mathrm{d}u_{k}}{\mathrm{d}\eta}=\dot{\imath}ku_{k}+\frac{\Phi'}{\Phi}v_{k}^{*} 
                 \hspace{0.5cm} \textrm{,} \hspace{0.5cm}
                 \frac{\mathrm{d}v_{k}}{\mathrm{d}\eta}=\dot{\imath}kv_{k}+\frac{\Phi'}{\Phi}u_{k}^{*} 
                 \hspace{0.5cm} \textrm{.}
              \end{equation} 
 \par
The equations (\ref{eq:solgeralcoefbogo}) are the Bogoliubov transformations \cite{birrel} of the gravitational waves in the Brans-Dicke theory and $u_{k}(\eta)$ and $v_{k}(\eta)$ are the Bogoliubov coefficients that satisfy the relation
              \begin{equation}
                 \left|u_{k}\right|^2-\left|v_{k}\right|^2=1 \hspace{0.3cm} \textrm{,}
              \end{equation}
result that guarantees the unity of the temporal evolution of these operators.
\par
The number operator is defined as $N_k = v_{k}^{*}v_{k}$. So, in our model the number of gravitons is given by            
\begin{equation}
N_k = \left|v_{k}\right|^2=\frac{|f_{k}|^2}{2k}\left(k^2+\left(\frac{\Phi'}{\Phi}\right)^2\right)+  \frac{1}{2k}\left|\frac{\mathrm{d}f_{k}}{\mathrm{d}\eta}\right|^2-            \frac{1}{2k}\frac{\Phi'}{\Phi}\left(f_{k}\frac{\mathrm{d}f^{*}_{k}}{\mathrm{d}\eta}+
f^{*}_{k}\frac{\mathrm{d}f_{k}}{\mathrm{d}\eta}\right) -\frac{1}{2}
\hspace{0.2cm} \textrm{,}
\end{equation}
where the relation between $\nu_k$ and $f_k$ is
\begin{equation}
                 v_{k}^{*}=\frac{\dot{\imath}}{\sqrt{2k}}\frac{\mathrm{d}f_{k}}{\mathrm{d}\eta}+
                           \frac{f_{k}}{\sqrt{2k}}\left(k-\dot{\imath}\frac{\Phi'}{\Phi}\right)
                           \hspace{0.2cm} \textrm{.}
\end{equation}
\par
When $k\eta>>1$ the number operator simplifies as
              \begin{equation}
                 N_k \approx k|f_{k}|^2\left(1+\frac{1}{2k^2}\left(\frac{\Phi'}{\Phi}\right)^2\right)-\frac{1}{2}
                 \hspace{0.3cm} \textrm{.}
              \end{equation}
\par
If $k\eta<<1$ we have
              \begin{equation}
                 N_k \approx k|f_{k}|^2-\frac{1}{2}
                 \hspace{0.3cm} \textrm{.}
              \end{equation}
\par
\begin{figure}[!t]
\begin{minipage}[t]{0.48\linewidth}
\includegraphics[width=\linewidth]{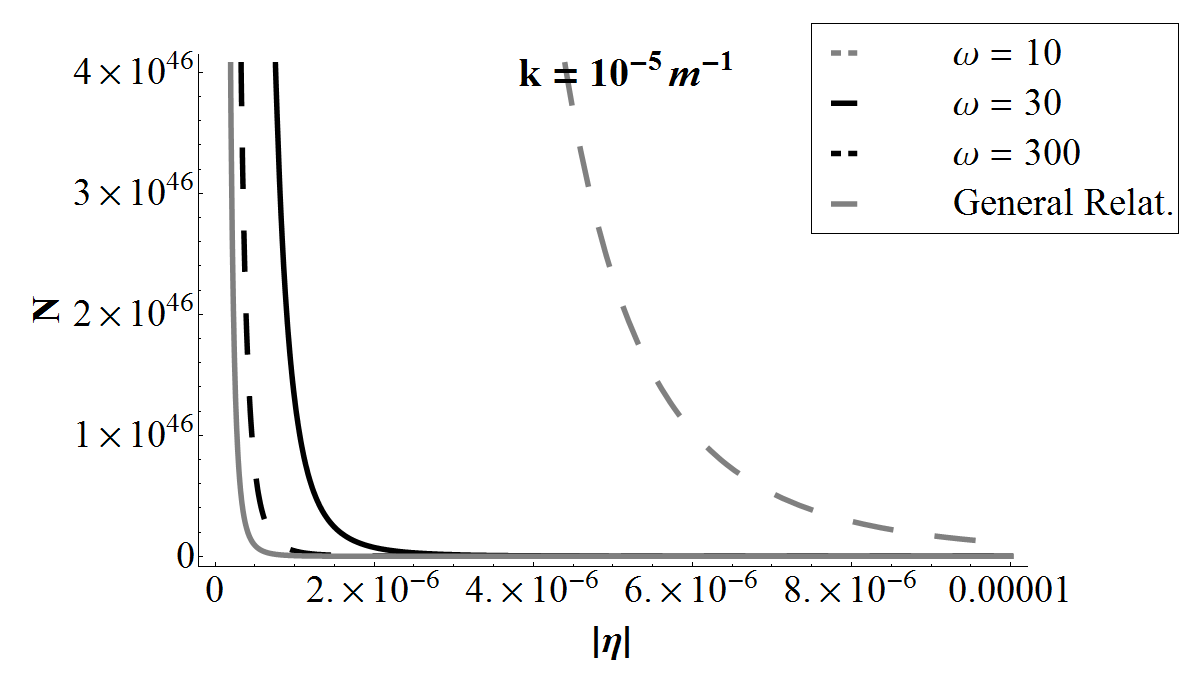}
\end{minipage} \hfill
\begin{minipage}[t]{0.48\linewidth}
\includegraphics[width=\linewidth]{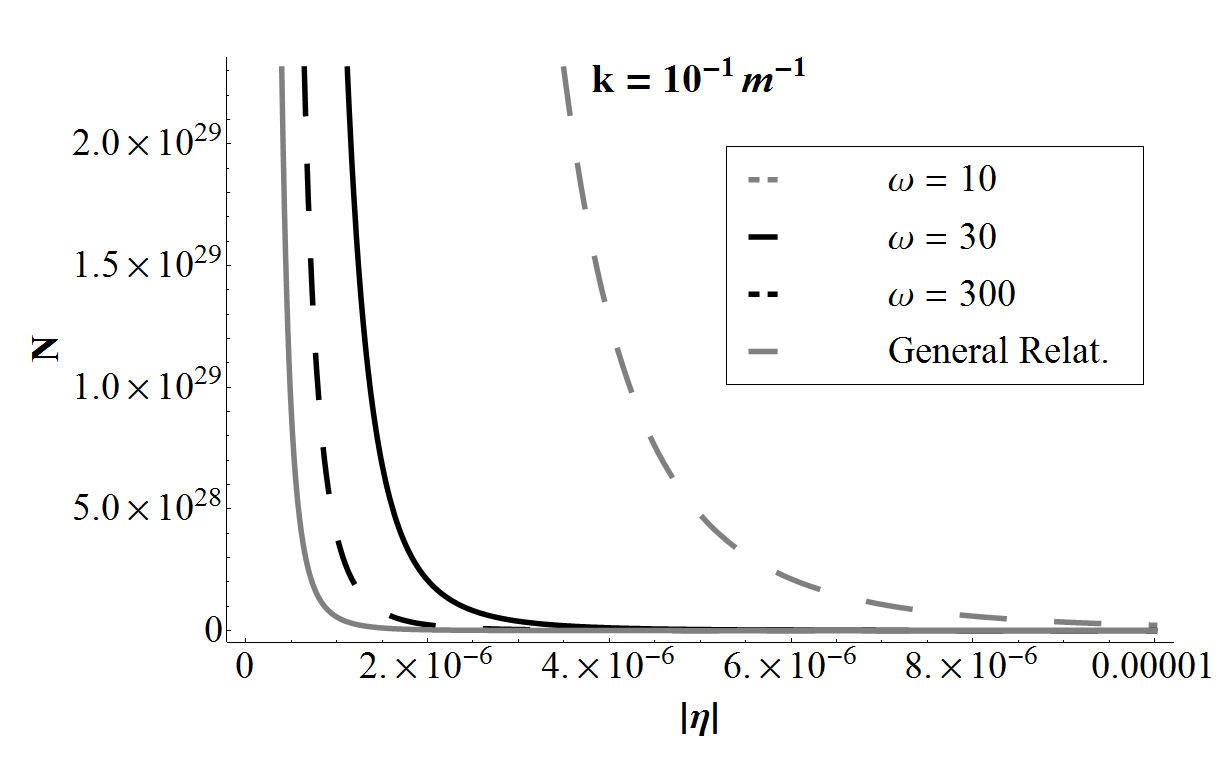}
\end{minipage}
\caption[Numero de particulas em funcao do tempo]{The number of gravitons produced by inflation in the Brans-Dicke theory for different values of $\omega$, and also in the General Relativity, for the wavenumbers $k=10^{-5}~m^{-1}$ and $k=10^{-1}~m^{-1}$.}
\label{fig01}
\end{figure}
\begin{figure}[!t]
\begin{minipage}[t]{0.48\linewidth}
\includegraphics[width=\linewidth]{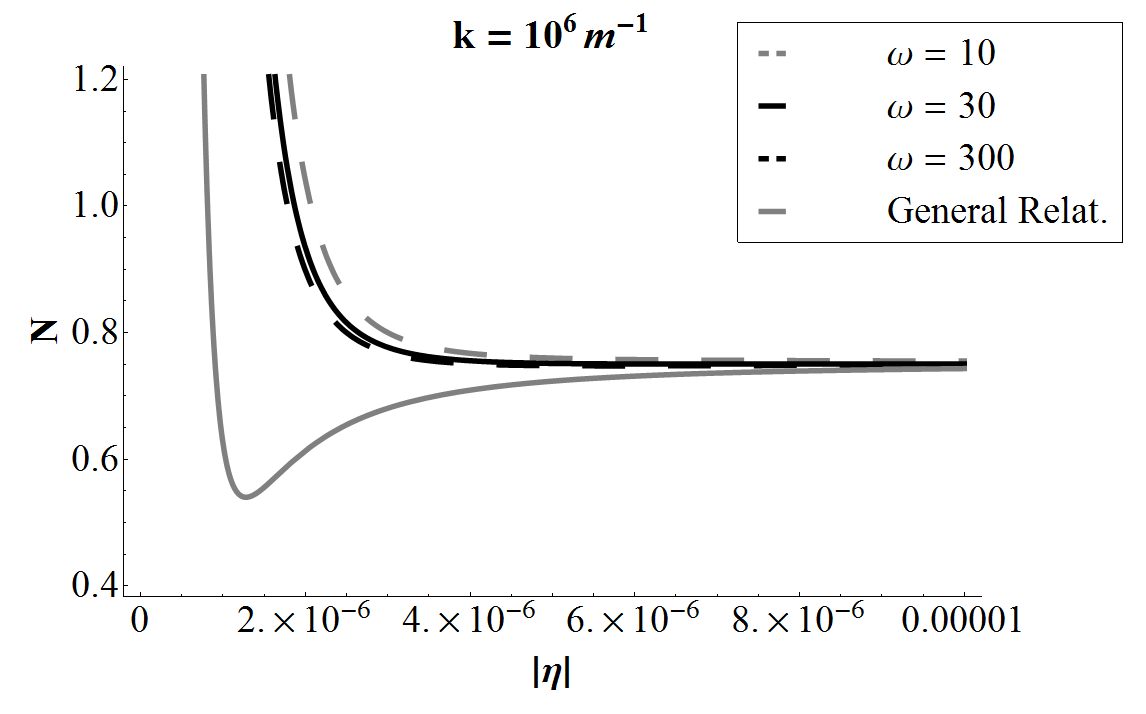}
\end{minipage} \hfill
\begin{minipage}[t]{0.48\linewidth}
\includegraphics[width=\linewidth]{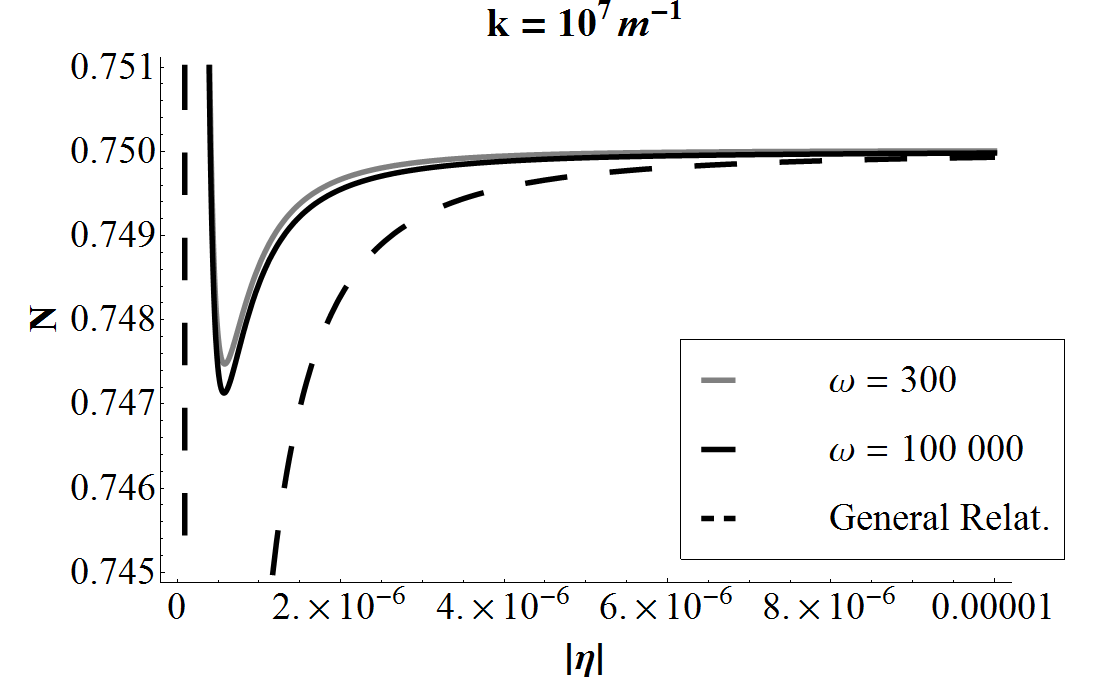}
\end{minipage}
\caption[Numero de particulas em funcao do tempo]{Same as the previous figure, but for wavenumbers $k=10^{6}~m^{-1}$ e $k=10^{7}~m^{-1}$.}
\label{fig02}
\end{figure}
\begin{figure}[!t]
\begin{minipage}[t]{0.48\linewidth}
\includegraphics[width=\linewidth]{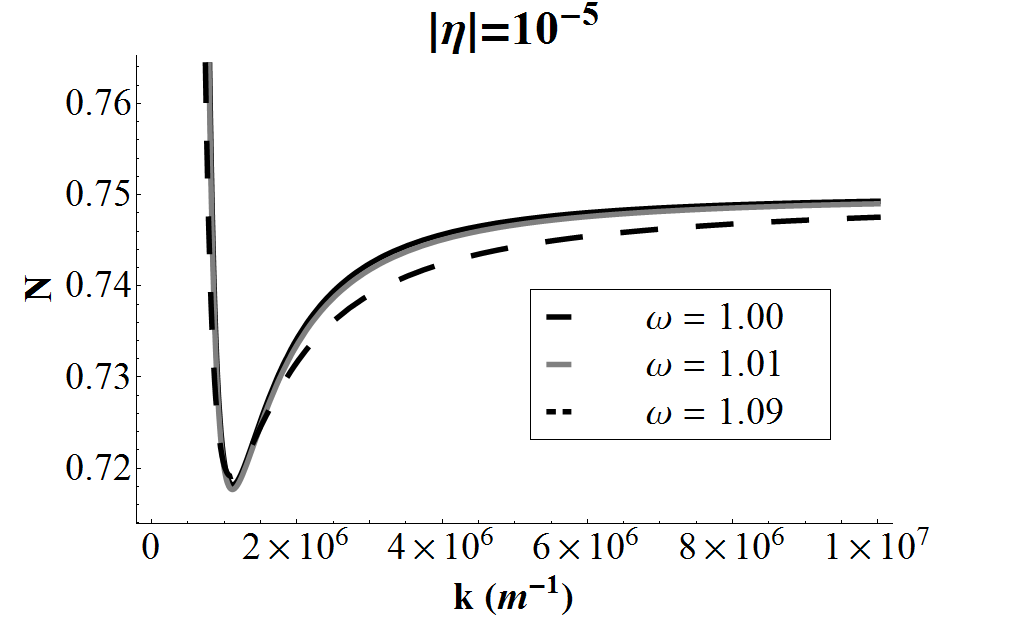}
\end{minipage} \hfill
\begin{minipage}[t]{0.48\linewidth}
\includegraphics[width=\linewidth]{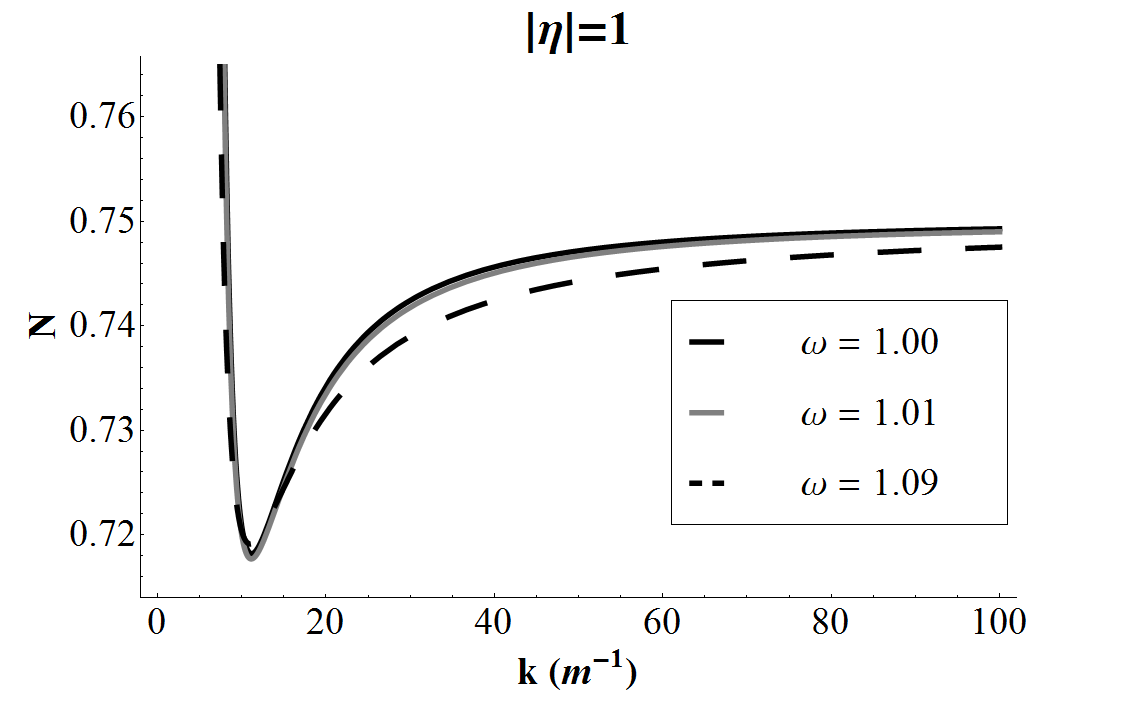}
\end{minipage}
\caption[Numero de particulas em funcao de $k$]{The number of gravitons for small values of $\omega$ at two different comoving times.}
\label{fig03}
\end{figure}
\begin{figure}[!t]
\begin{minipage}[t]{0.48\linewidth}
\includegraphics[width=\linewidth]{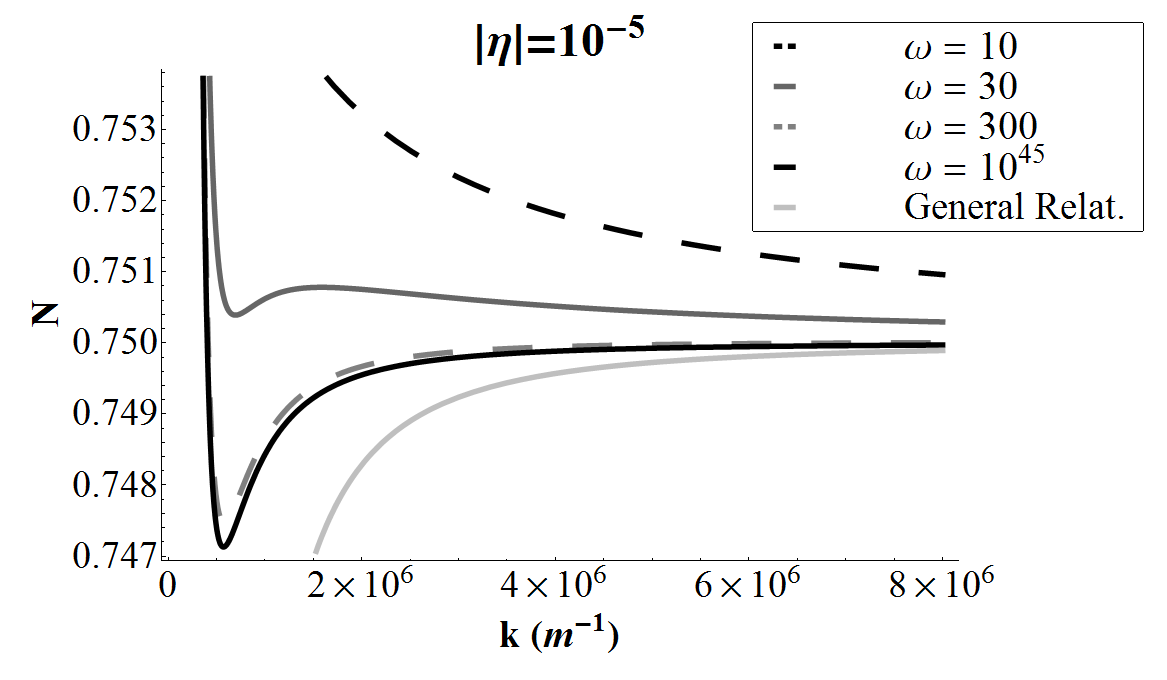}
\end{minipage} \hfill
\begin{minipage}[t]{0.48\linewidth}
\includegraphics[width=\linewidth]{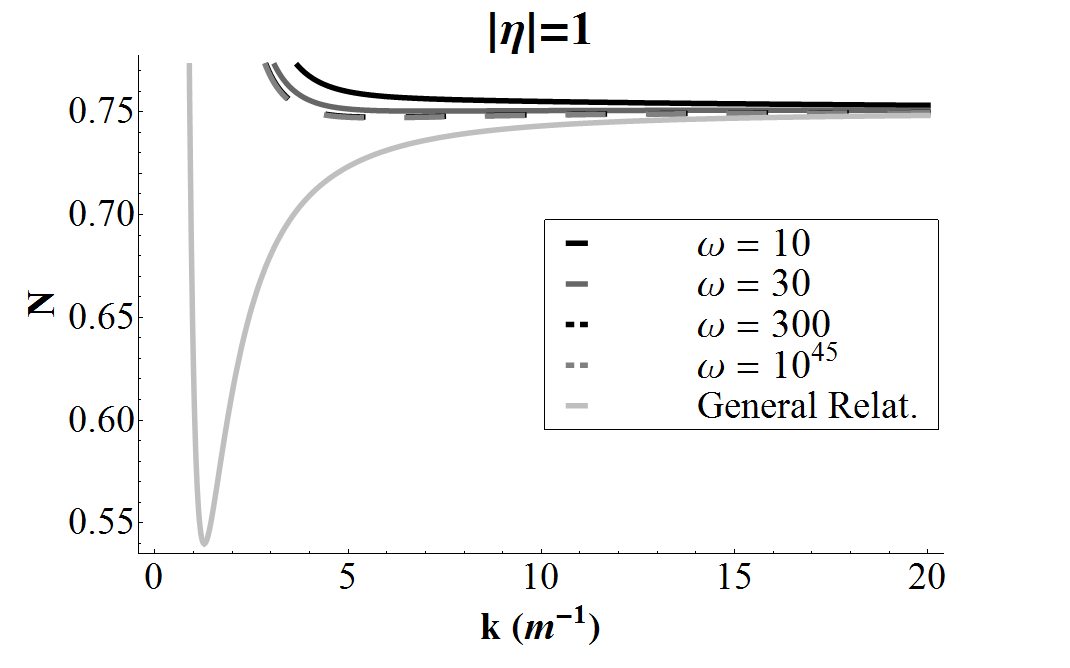}
\end{minipage}
\caption[Numero de particulas em funcao de $k$]{The number of gravitons for large values of $\omega$ and for General Relativity.}
\label{fig04}
\end{figure}

In Fig.~\ref{fig01}~-~Fig.~\ref{fig04} we show the behavior of the number of particles in the inflation era in the Brans-Dicke theory to different values of $\omega$ and small and big values of the wave number and we compare with the General Relativity. The aim here is, first of all, see how the Brans-Dicke model behaves for different values of $\omega$, since local observations show that $\omega$ is big and, as we have written before, small values should be more interesting in large scales. Secondly, when gravitational waves of cosmological origin are observed, the Brans-Dicke theory and General Relativity may be confronted regarding the evolution of the tensor modes.

\section{Observables in the Brans-Dicke theory}
\label{section_spectrum}
\par
In this section we apply the formulation of quantum gravitational waves in Brans-Dicke theory to determine the most common and interesting observables  in cosmology. The experimental data at present do not allow us to include or exclude the presence of a primordial spectrum of relic gravitons compatible with any model. The idea is that these theoretical results calculated here might in the near future be compared with the observational data obtained by the projects related to the search for gravitational waves of cosmological origin.

\subsection{Power Spectrum}

\par
The power spectrum is related by the Fourier transform of the two-point correlation function of the tensor modes in the following way \cite{inflacao}
\begin{equation}
                 \label{eq:funccorrelacao}
                 \left\langle0\right|\hat{h}_{ij}^{\dagger}(\vec{x},\eta)\hat{h}^{ij}(\vec{y},\eta)\left|0\right\rangle=
                 \int{P_{T}(k,\eta)\frac{\sin{(kr)}}{kr} \mathrm{d}(\ln{k})}
                 \hspace{0.2cm} \textrm{,}
\end{equation}
where $P_{T}(k,\eta)$ is the power spectrum and the state $\left|0\right\rangle$ is annihilated by $\hat{a}_{\vec{k}}$ .
         \par   
The process of quantization of gravitational waves makes the classical field $h_{ij}$ to be transformed into an operator $\hat{h}_{ij}$. This operator is given by the equation (\ref{oper01}) where the vacuum state $\left|0\right\rangle$, in which there are no particles, is annihilated by the annihilation operator
              \begin{equation}
                 \hat{a}_{\vec{k}}\left|0\right\rangle=0
                 \hspace{0.3cm} \textrm{.}
              \end{equation}  
         \par
The calculations of the expectation value of the field operators yields
\begin{equation}
                 \left\langle0\right|\hat{h}_{ij}^{\dagger}(\vec{x},\eta)\hat{h}^{ij}(\vec{y},\eta)\left|0\right\rangle=
                    64\pi\int{\frac{\mathrm{d}^3k}{(2\pi)^{3}} \frac{\left|f_{k}(\eta)\right|^2}{a^2\phi}
                     e^{-\dot{\imath}\vec{k}\cdot(\vec{x}-\vec{y})} } 
                     \hspace{0.2cm} \textrm{.}
\end{equation}
         \par
Knowing that gravitons can be produced in isotropic models and using spherical coordinates, we have
\begin{equation}
                 \left\langle0\right|\hat{h}_{ij}^{\dagger}(\vec{x},\eta)\hat{h}^{ij}(\vec{y},\eta)\left|0\right\rangle=
                    64\pi\int{\frac{k^2}{(2\pi)^{2}} \frac{\left|f_{k}\right|^2}{a^2\phi} 
                      \frac{2\sin{(kr)}}{kr} \mathrm{d}k}
                      \hspace{0.2cm} \textrm{,}
\end{equation}
\par
Comparing the equation (\ref{eq:funccorrelacao}) with the previous equation we have, within the Brans-Dicke model, the power spectrum of the tensor modes parametrized as
              \begin{equation}
                 P_{T}(k,\eta)=32\pi\frac{k^{3}}{\pi^2}\frac{\left|f_{k}\right|^2}{a^2\phi}
                 \hspace{0.3cm} \textrm{.}
              \end{equation}  

\begin{figure}[!t]
\begin{minipage}[t]{0.48\linewidth}
\includegraphics[width=\linewidth]{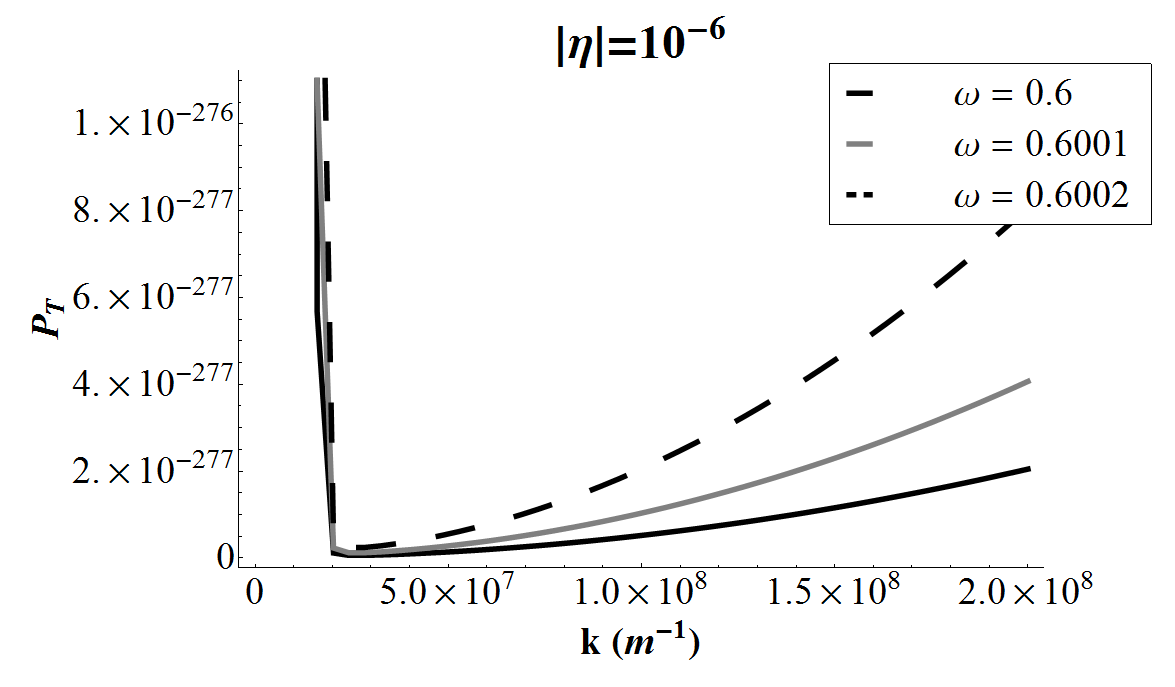}
\end{minipage} \hfill
\begin{minipage}[t]{0.48\linewidth}
\includegraphics[width=\linewidth]{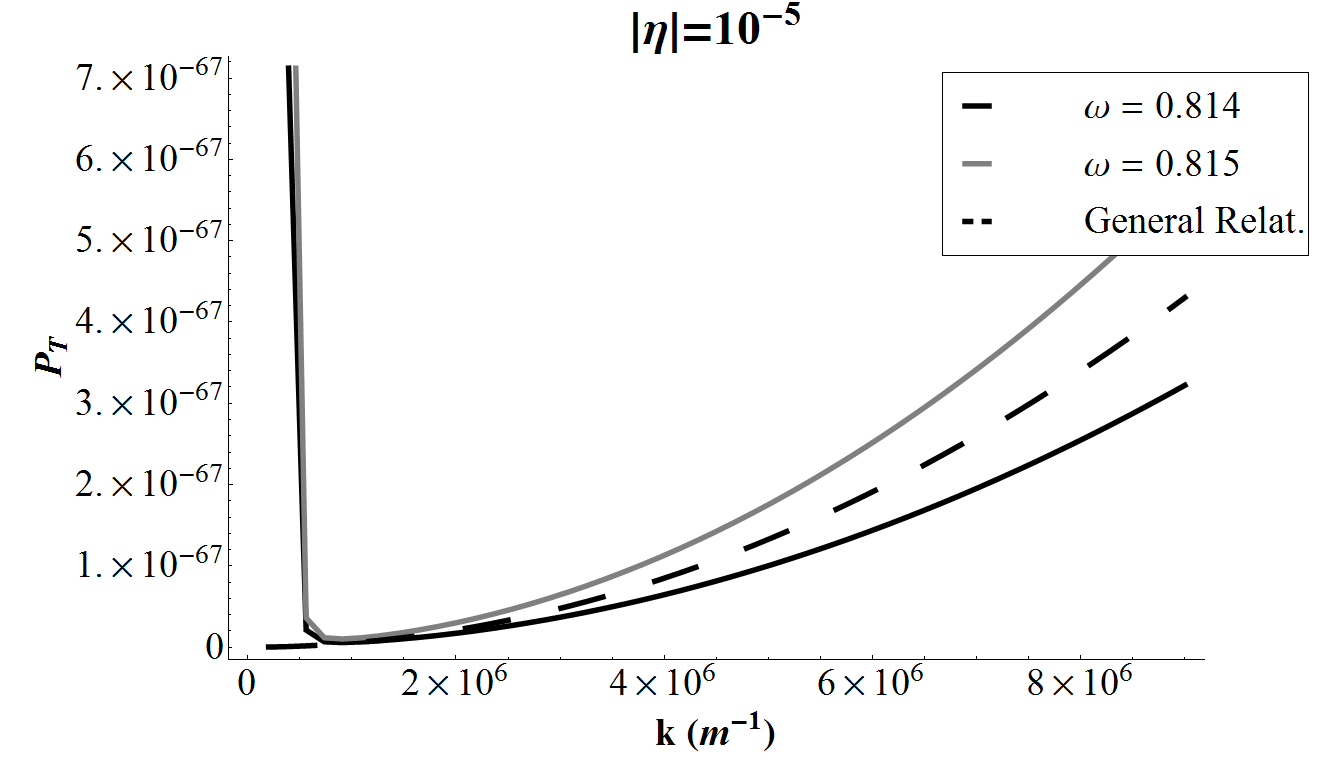}
\end{minipage}
\caption[Espectro de potencia em funcao de $k$]{The power spectrum of the gravitational waves in the Brans-Dicke theory as a function of wavenumber for small values of $\omega$.}
\label{fig:potencia1}
\end{figure}

\begin{figure}[!t]
\begin{minipage}[t]{0.48\linewidth}
\includegraphics[width=\linewidth]{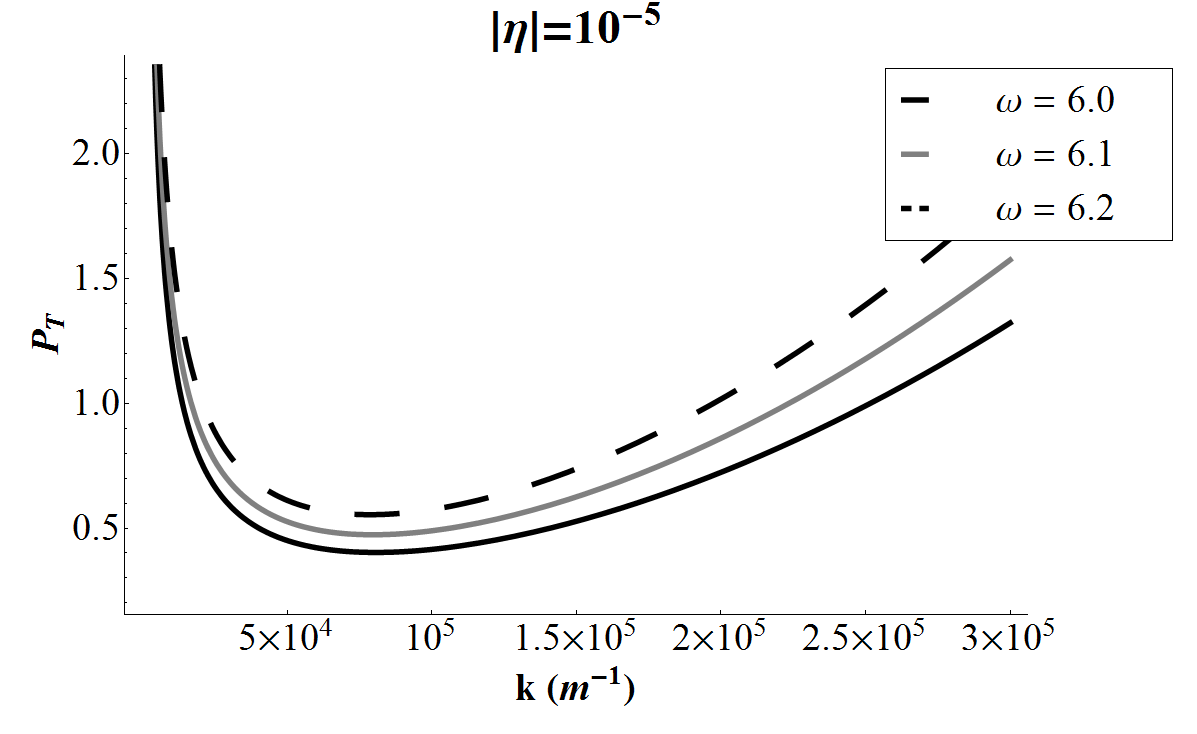}
\end{minipage} \hfill
\begin{minipage}[t]{0.48\linewidth}
\includegraphics[width=\linewidth]{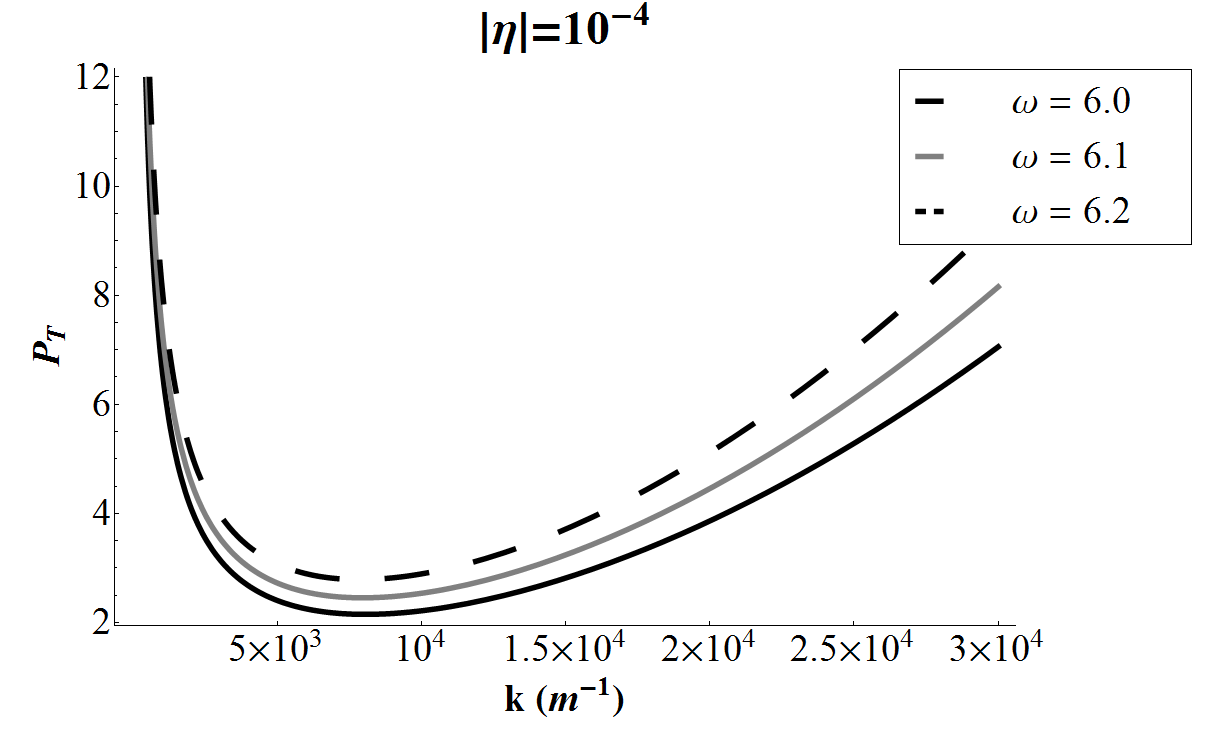}
\end{minipage}
\caption[Espectro de potencia em funcao de $k$]{The power spectrum of the gravitational waves in the Brans-Dicke theory for large values of
$\omega$.}
            \label{fig:potencia5}
\end{figure}      
         \par
In the inflation phase the function $f_{k}(k|\eta|)$ is defined as
              \begin{equation}
                 f_{k}(k|\eta|)=\frac{\sqrt{\pi}}{2}\sqrt{|\eta|}\mathcal{H}_{\nu}^{(2)}(k|\eta|)\quad,
              \end{equation}
and by using the small argument limit of the Hankel functions \cite{abramowitz}, we have
              \begin{equation}
                 P_{T}(k,\eta)=\frac{8\pi2^{2\nu}\Gamma^{2}(\nu)}{\pi^3}\frac{k^2}{a^2\phi}(k|\eta|)^{1-2v}
                 \hspace{0.3cm} \textrm{.}
              \end{equation}
         \par 
By definition $H=\frac{\dot{a}}{a}=\frac{1}{a^2}\frac{\mathrm{d}a}{\mathrm{d}\eta}$, and the conformal time in this model can be written as
              \begin{equation}
                 \label{eq:etaBD}
                 |\eta|=\left(\frac{2\omega+1}{2\omega-1} \right)\frac{1}{aH}
                 \hspace{0.2cm} \textrm{.}
              \end{equation}
Thus, the power spectrum in the Brans-Dicke theory during inflation for the modes out of the horizon becomes
\begin{equation}
                 P_{T}(k,\eta)=
                \frac{8\pi2^{2\nu}\Gamma^{2}(\nu)}{\pi^3}\left(\frac{2\omega+1}{2\omega-1} \right)^{1-2\nu}
                 \frac{H^2}{\phi}\left(\frac{k}{aH}\right)^{3-2\nu}
                 \hspace{0.2cm} \textrm{.}
\end{equation}
\par
In Fig.~\ref{fig:potencia1} - Fig.~\ref{fig:potencia5} we show the behavior of the power spectrum $P_T(k, \eta)$ as a function of the wavenumber and of the parameter $\omega$.
      
\subsection{Spectral Index}
      
         \par
The spectral index is obtained by the expression
              \begin{equation}
                 n_{T}=\frac{\mathrm{d}{\ln{P_{T}(k,\eta)}}}{\mathrm{d}\ln{k}}\Bigg|_{\alpha} 
                 \hspace{0.2cm} \textrm{,}
              \end{equation}     
where $\alpha\equiv aH=\left(\frac{2\omega+1}{2\omega-1}\right)k$ is the moment when the tensorial modes cross the horizon again. For this calculation we introduce the parameter $\epsilon$, that measures the rate of decrease of the Hubble parameter during the inflation phase and it is given by
               \begin{equation}
                  \label{eq:epsilon}
                  \epsilon\equiv \frac{\mathrm{d}}{\mathrm{d}t}\left(\frac{1}{H}\right)=
                       -\frac{1}{aH^2}\frac{\mathrm{d}H}{\mathrm{d}\eta}
                       \hspace{0.2cm} \textrm{.}
               \end{equation}
          \par    
In the Brans-Dicke theory we obtain that
               \begin{equation}
                  \label{eq:epsilonBD}
                  \epsilon=\frac{2}{2\omega+1}
                  \hspace{0.2cm} \textrm{.}
               \end{equation}  
         \par   
Thus, we have
\begin{equation}
\label{espectral}
                  n_{T}=\frac{\mathrm{d}}{\mathrm{d}\ln{k}}\left[\ln{\left(\frac{H^2}{\phi}\right)}\right]
                  \Bigg|_{\alpha}=
                 \biggl(2\frac{k}{H}\frac{\mathrm{d}H}{\mathrm{d}k}
                  -\frac{\mathrm{d}\ln{\phi}}{\mathrm{d}\ln{k}}\biggr)\Bigg|_{\alpha}
                  \hspace{0.2cm} \textrm{.}
\end{equation}
         \par
The first term on the right side is calculated as follows
\begin{equation}
                   \label{eq:Hetak}
                   2\frac{k}{H}\frac{\mathrm{d}H}{\mathrm{d}k}\Bigg|_{\alpha}=
                      2\frac{k}{H}\frac{\mathrm{d}H}{\mathrm{d}|\eta|}\frac{\mathrm{d}|\eta|}{\mathrm{d}k}
                      \Bigg|_{\alpha}=
                     2\frac{k}{H}\frac{\mathrm{d}H}{\mathrm{d}\eta}
                      \left(\frac{\mathrm{d}|\eta|}{\mathrm{d}\eta}\right)^{-1}\frac{\mathrm{d}|\eta|}{\mathrm{d}k}
                      \Bigg|_{\alpha}
                      \hspace{0.2cm} \textrm{,}
\end{equation}
and, by using the equation (\ref{eq:etaBD}), we have \cite{dodelson}
\begin{equation}
                  \label{eq:etak}
                  \frac{\mathrm{d}|\eta|}{\mathrm{d}k}\Bigg|_{\alpha}=
                     \frac{\mathrm{d}}{\mathrm{d}k}\left(\left(\frac{2\omega+1}{2\omega-1}\right)\frac{1}{aH}\right)
                     \Bigg|_{\alpha}=
                     \frac{\mathrm{d}}{\mathrm{d}k}\left(\frac{1}{k}\right)=-\frac{1}{k^{2}}
                     \hspace{0.2cm} \textrm{.}
\end{equation}
\par 
In this way, with the help of the equations (\ref{eq:epsilonBD}), (\ref{eq:Hetak}) and (\ref{eq:etak}), the equation (\ref{espectral}) can be written as
               \begin{equation}
                  \label{eq:indiceomegaphi}
                  n_{T}=-2\epsilon\left(\frac{2\omega+1}{2\omega-1}\right)-\frac{\mathrm{d}\ln{\phi}}{\mathrm{d}\ln{k}}
                  \Bigg|_{\alpha}
                  \hspace{0.2cm} \textrm{.}
               \end{equation}
\par
When $\omega\rightarrow\infty$ and $\phi$ is constant we obtain the same result that in General Relativity \cite{dodelson}, \textit{i. e.}, $n_T = - 2\epsilon$.
         \par  
Likewise, we can calculate the term dependent on the scalar field $\phi$ in equation (\ref{eq:indiceomegaphi})
\begin{equation}
                  \frac{\mathrm{d}\ln{\phi}}{\mathrm{d}\ln{k}}\Bigg|_{\alpha}=
                  \frac{k}{\phi}\frac{\mathrm{d}\phi}{\mathrm{d}k}\Bigg|_{\alpha}=
                  \frac{k}{\phi}\frac{\mathrm{d}\phi}{\mathrm{d}\eta}\frac{\mathrm{d}\eta}{\mathrm{d}k}
                  \Bigg|_{\alpha}=
                  \frac{k}{\phi}\frac{\mathrm{d}\phi}{\mathrm{d}\eta}\left(\frac{\mathrm{d}|\eta|}{\mathrm{d}\eta}\right)^{-1}
                  \frac{\mathrm{d}|\eta|}{\mathrm{d}k}\Bigg|_{\alpha}  
                  \hspace{0.2cm} \textrm{,}
\end{equation}
that with the help of equation (\ref{eq:etaBD}) and with the background solutions of the scalar field we obtain the following result for the spectral index $n_T$
               \begin{equation}
                  n_{T}=-2\epsilon\left(\frac{2\omega+1}{2\omega-1}\right)-\frac{4}{2\omega-1}
                  \hspace{0.2cm} \textrm{,}
               \end{equation}   
Again, the above result gives the same value of the spectral index of the tensorial modes of General Relativity at the time of inflation, when $\omega
\rightarrow \infty $.
\par
We can write the expression of $n_T$ in terms of the parameter $\omega$. For that we replace the relation (\ref{eq:epsilonBD}) in the result found above, such that
               \begin{equation}
                  \label{eq:indiceomega}
                  n_{T}=\frac{-8}{2\omega-1} \hspace{0.2cm} \textrm{.}
               \end{equation}
               
                              
         \par
By using (\ref{eq:epsilon}) we obtain
               \begin{equation}
                  2\omega=\frac{2}{\epsilon}-1
                  \hspace{0.2cm} \textrm{,}
               \end{equation}      
which can be replaced in the final result for the spectral index such that
               \begin{equation}
                  n_{T}=\frac{4\epsilon}{\epsilon-1}=-4\epsilon(1-\epsilon)^{-1}
                  \hspace{0.2cm} \textrm{.}
               \end{equation}
For small values of $\epsilon$ we can expand the term $(1-\epsilon)^{-1}$ obtaining
               \begin{equation}
                  n_{T}= -4\epsilon(1+\epsilon+\mathcal{O}(\epsilon^2))
                  \hspace{0.2cm} \textrm{,}
               \end{equation}
and considering only zero-order terms we have
               \begin{equation}
                  n_{T}\approx-4\epsilon
                  \hspace{0.2cm} \textrm{.}
               \end{equation}

\subsection{Energy Density}
      
         \par
With the Lagrangian density of the gravitational waves in the Brans-Dicke theory, equation (\ref{lagran01}), we obtain the energy-momentum tensor of the gravitatinal waves
               \begin{equation}
                  {T}_{\mu\nu}=-\frac{\phi}{8\pi}\frac{1}{4}
                  \left(\partial_{\mu}h_{ij}\partial_{\nu}h^{ij}-
                  \frac{1}{2}\bar{g}_{\mu\nu}\bar{g}^{\alpha\beta}\partial_{\alpha}h_{ij}\partial_{\beta}h^{ij}\right)
                  \hspace{0.2cm} \textrm{.}
               \end{equation}  
         \par
Considering 
               \begin{equation}
                  h_{ij}=\sqrt{16\pi}\sum_{\lambda = \otimes , \oplus}{^{(\lambda)}{\epsilon_{ij}}h^{(\lambda)}}
                  \hspace{0.3cm} \textrm{,} \hspace{0.2cm}
                  h^{(\lambda)}=\frac{\mu^{(\lambda)}}{a\sqrt{\phi}}
                  \hspace{0.2cm} \textrm{,}
               \end{equation}
we find that
               \begin{equation}
                  {T}_{\mu\nu}=-2\phi\left(\partial_{\mu}h\partial_{\nu}h-
                     \frac{1}{2}\bar{g}_{\mu\nu}\bar{g}^{\alpha\beta}\partial_{\alpha}h\partial_{\beta}h\right)
                     \hspace{0.2cm} \textrm{,}
               \end{equation}
where
               \begin{equation}
                  h=h^{\otimes}=h^{\oplus}
                  \hspace{0.2cm} \textrm{.}
               \end{equation}
         \par
The energy density is the (00) component of the energy-momentum tensor of the gravitational waves
               \begin{equation}
                  \rho={T}_{0}^{~0}=\phi\left(\frac{h'^{2}}{a^2}+
                       \bar{g}^{ij}\partial_{i}h\partial_{j}h\right)
                  \hspace{0.2cm} \textrm{.}
               \end{equation}
         \par
To calculate the quantum energy density we first substitute the following
               \begin{equation}
                  h\rightarrow\hat{h}=\frac{\hat{\mu}}{a\sqrt{\phi}}
                  \hspace{0.2cm} \textrm{,}
               \end{equation}
such that the expectation value of the energy density is given by
               \begin{equation}
                  \left\langle\rho\right\rangle=\left\langle0\right|\rho\left|0\right\rangle=
                  \frac{\phi}{a^{2}}\left(\left\langle0\right|\hat{h}'\hat{h}'^{*}\left|0\right\rangle+
                  \left\langle0\right|\nabla\hat{h}\cdot\nabla\hat{h}^{*}\left|0\right\rangle\right)
                  \hspace{0.2cm} \textrm{.}
               \end{equation}
         \par
With the help of the expansions
\begin{eqnarray}
                 \label{eq:operadores}
                 \hat{\mu}=\frac{1}{2}\int{\frac{\mathrm{d}^{3}k}{(2\pi)^{3/2}}\left[\hat{\mu}_{\vec{k}}(\eta)e^{-\dot{\imath}\vec{k}\cdot\vec{x}} 
                 +\hat{\mu}_{\vec{k}}^{\dagger}(\eta)e^{\dot{\imath}\vec{k}\cdot\vec{x}} \right]} \quad, \nonumber \\
                 \hat{\pi}=\frac{1}{2}\int{\frac{\mathrm{d}^{3}k}{(2\pi)^{3/2}}\left[\hat{\pi}_{\vec{k}}(\eta)e^{-\dot{\imath}\vec{k}\cdot\vec{y}} 
                 +\hat{\pi}_{\vec{k}}^{\dagger}(\eta)e^{\dot{\imath}\vec{k}\cdot\vec{y}} \right]} \hspace{0.2cm} \textrm{,}
\end{eqnarray}
the solutions
\begin{eqnarray} 
                 \label{eq:solgeloperadores}
                 \hat{\mu}_{\vec{k}}=\hat{a}_{\vec{k}}(\eta_0)f_{k}(\eta)+\hat{a}_{-\vec{k}}^{\dagger}(\eta_0)f_{k}^{*}(\eta) \quad, \nonumber \\
                 \hat{\pi}_{\vec{k}}=\hat{a}_{\vec{k}}(\eta_0)g_{k}(\eta)+\hat{a}_{-\vec{k}}^{\dagger}(\eta_0)g_{k}^{*}(\eta)
\hspace{0.2cm} \textrm{,}
\end{eqnarray}
and the commutation relations
\begin{equation}
                 \label{eq:comutacao1}
                 \left[\hat{a}_{\vec{k}},\hat{a}_{\vec{q}}^{\dagger} \right]=\delta^{(3)}(\vec{k}-\vec{q})   \hspace{0.2cm} \textrm{,}
\end{equation}
we obtain
\begin{eqnarray}
                  & &\left\langle0\right|\hat{h}'\hat{h}'^{*}\left|0\right\rangle=
                    \frac{1}{a^{2}\phi}\int \frac{\mathrm{d}^{3}k}{(2\pi)^{3}}\biggl\{|g_{k}|^{2}
                    +\left(\frac{\Phi'}{\Phi}\right)^{2}|f_{k}|^{2}-
                    \left(\frac{\Phi'}{\Phi}\right)(g_{k}f_{k}^{*}+g_{k}^{*}f_{k})
                    \biggr\}e^{-\dot{\imath}\vec{k}\cdot(\vec{x}-\vec{y})}
                  \\
                  & &\left\langle0\right|\nabla\hat{h}\cdot\nabla\hat{h}^{*}\left|0\right\rangle=
                   \frac{1}{a^{2}\phi}\int{\frac{\mathrm{d}^{3}k}{(2\pi)^{3}} k^{2}|f_{k}|^{2}e^{-\dot{\imath}\vec{k}\cdot(\vec{x}-\vec{y})}}
                    \hspace{0.2cm} \textrm{.}
\end{eqnarray}
Thus, the expectation value of the energy density of the gravitational waves in the Brans-Dicke theory can be write as
\begin{equation}
                 \left\langle\rho\right\rangle=
                 \frac{1}{a^4}\int \frac{\mathrm{d}^{3}k}{(2\pi)^{3}}\biggl[
                       |g_{k}|^{2}+\left(k^{2}+\left(\frac{\Phi'}{\Phi}\right)^{2}\right)|f_{k}|^{2}-
                       \frac{\Phi'}{\Phi}\left(g_{k}f_{k}^{*}+g_{k}^{*}f_{k}\right)\biggr]e^{-\dot{\imath}\vec{k}(\vec{x}-\vec{y})}
                       \hspace{0.2cm} \textrm{.}
\end{equation}     
\par
Considering the presence of an isotropic background of relic gravitons we obtain
\begin{equation}
                 \left\langle\rho\right\rangle=
                 \frac{1}{a^4}\int \frac{k^{2}\mathrm{d}k}{(2\pi)^{3}}\biggl[
                       |g_{k}|^{2}+\left(k^{2}+\left(\frac{\Phi'}{\Phi}\right)^{2}\right)|f_{k}|^{2}-
                    \frac{\Phi'}{\Phi}\left(g_{k}f_{k}^{*}+g_{k}^{*}f_{k}\right)\biggr]\frac{2\sin{kr}}{kr}
                       \hspace{0.2cm} \textrm{,}
\end{equation} 
where $r=|\vec{x}-\vec{y}|$. In the limit $\vec{x}\rightarrow\vec{y}$ we have
\begin{equation}
                  \left\langle\rho\right\rangle=
                  \frac{1}{a^4}\int \frac{k^{2}\mathrm{d}k}{2\pi^{2}}\biggl[
                       |g_{k}|^{2}+\left(k^{2}+\left(\frac{\Phi'}{\Phi}\right)^{2}\right)|f_{k}|^{2}-
                       \frac{\Phi'}{\Phi}\left(g_{k}f_{k}^{*}+g_{k}^{*}f_{k}\right)\biggr]
                       \hspace{0.2cm} \textrm{.}
\end{equation}
         \par
The energy density of the relic gravitons per logarithmic interval of the wavenumber will be given by
\begin{equation}
                  \frac{\mathrm{d}\left\langle\rho\right\rangle}{\mathrm{d}\ln{k}}=
                  \frac{k^{3}}{2\pi^{2}a^{4}}\biggl[|g_{k}|^{2}+\left(k^{2}+\left(\frac{\Phi'}{\Phi}\right)^{2}\right)|f_{k}|^{2}-
                  \frac{\Phi'}{\Phi}\left(g_{k}f_{k}^{*}+f_{k}g_{k}^{*}\right)\biggr]
                  \hspace{0.2cm} \textrm{.}
\end{equation}
\par
The spectral energy density per logarithmic interval of the wavenumber  is defined as
               \begin{equation}
                  \Omega(k,\eta)=\frac{1}{\rho_{c}}\frac{\mathrm{d}\left\langle\rho\right\rangle}{\mathrm{d}\ln{k}}
                  \hspace{0.2cm} \textrm{,}
               \end{equation}
\textit{i. e.},
\begin{equation}
                  \Omega(k,\eta)=\frac{8\pi\phi^{-1}}{3H_{0}^{2}}\frac{k^{3}}{2\pi^{2}a^{4}}
                  \biggl[|g_{k}|^{2}+ \left(k^{2}+\left(\frac{\Phi'}{\Phi}\right)^{2}\right)|f_{k}|^{2}-
                  \frac{\Phi'}{\Phi}\left(g_{k}f_{k}^{*}+f_{k}g_{k}^{*}\right)\biggr]
                  \hspace{0.2cm} \textrm{,}
\end{equation}
where $\rho_c$ is the critical density of the Universe.                         
\par
For all the oscillation modes that are inside the Hubble radius (\textit{i. e.} $k\eta >> 1$), we have $g_ {k}\approx\mp\dot\imath kf_ {k}$. This implies that
               \begin{equation}
                  \Omega(k,\eta)\approx\frac{8\pi k^5}{3\pi^2H_{0}^2\phi a^4}|f_{k}|^2
                  \left(1+\frac{1}{2k^2}\left(\frac{\Phi'}{\Phi}\right)^2\right)
                  \hspace{0.2cm} \textrm{,}
               \end{equation}

and, rewriting in terms of the power spectrum we have
               \begin{equation}
                  \Omega(k,\eta)\approx\frac{k^2}{12H_{0}^2a^2}P_{T}\left(1+\frac{1}{2k^2}\left(\frac{\Phi'}{\Phi}\right)^2\right)
                  \hspace{0.2cm} \textrm{.}
               \end{equation}
          \par
For modes outside the Hubble radius (\textit{i. e.} $k\eta<<1$) we have $g_{k}\approx-\frac{\Phi'}{\Phi}f_{k}$. This result allow us to find that
               \begin{equation}
                  \Omega(k,\eta)\approx \frac{8\pi k^5}{6\pi^2 H_{0}^2\phi a^4}|f_{k}|^2
                  \hspace{0.2cm} \textrm{,}
               \end{equation}
which can also be rewritten as a function of the power spectrum $P_T$
               \begin{equation}
                  \Omega(k,\eta)\approx\frac{k^2}{24H_{0}^2a^2}P_{T}(k,\eta)
                  \hspace{0.2cm} \textrm{.}
               \end{equation}
\par
\begin{figure}[!t]
\begin{minipage}[t]{0.48\linewidth}
\includegraphics[width=\linewidth]{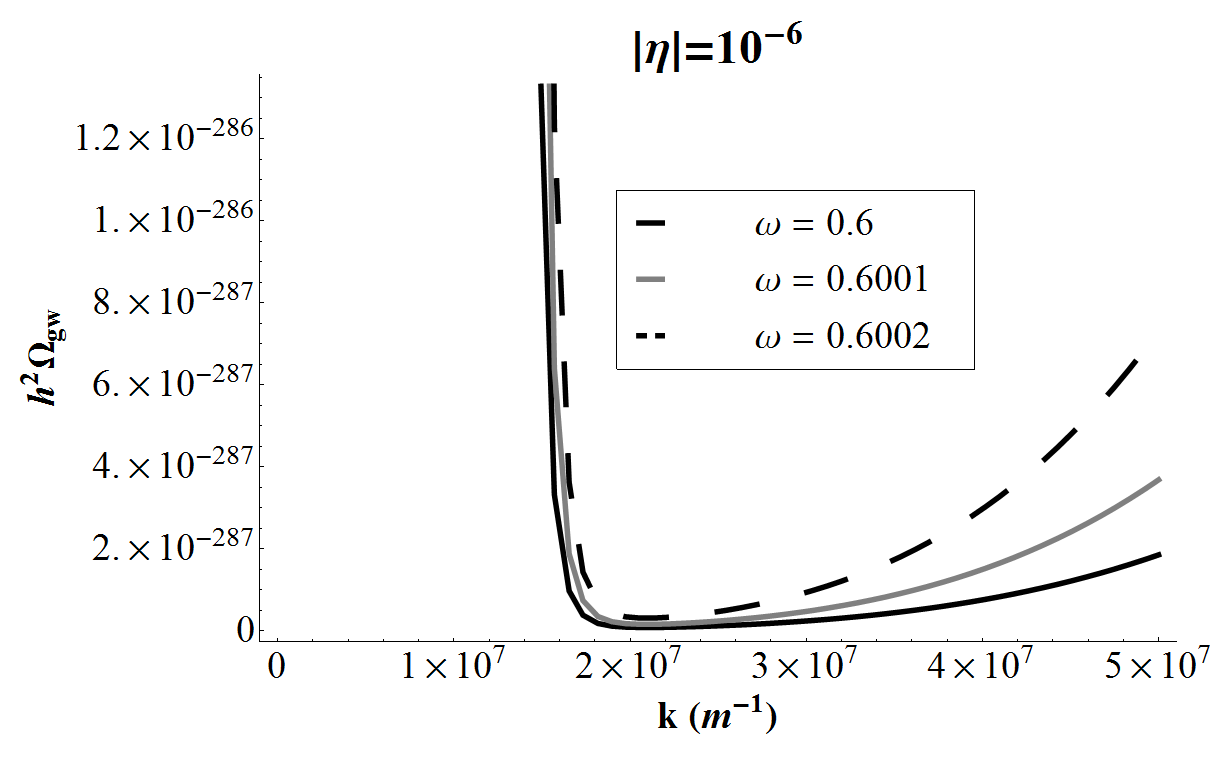}
\end{minipage} \hfill
\begin{minipage}[t]{0.48\linewidth}
\includegraphics[width=\linewidth]{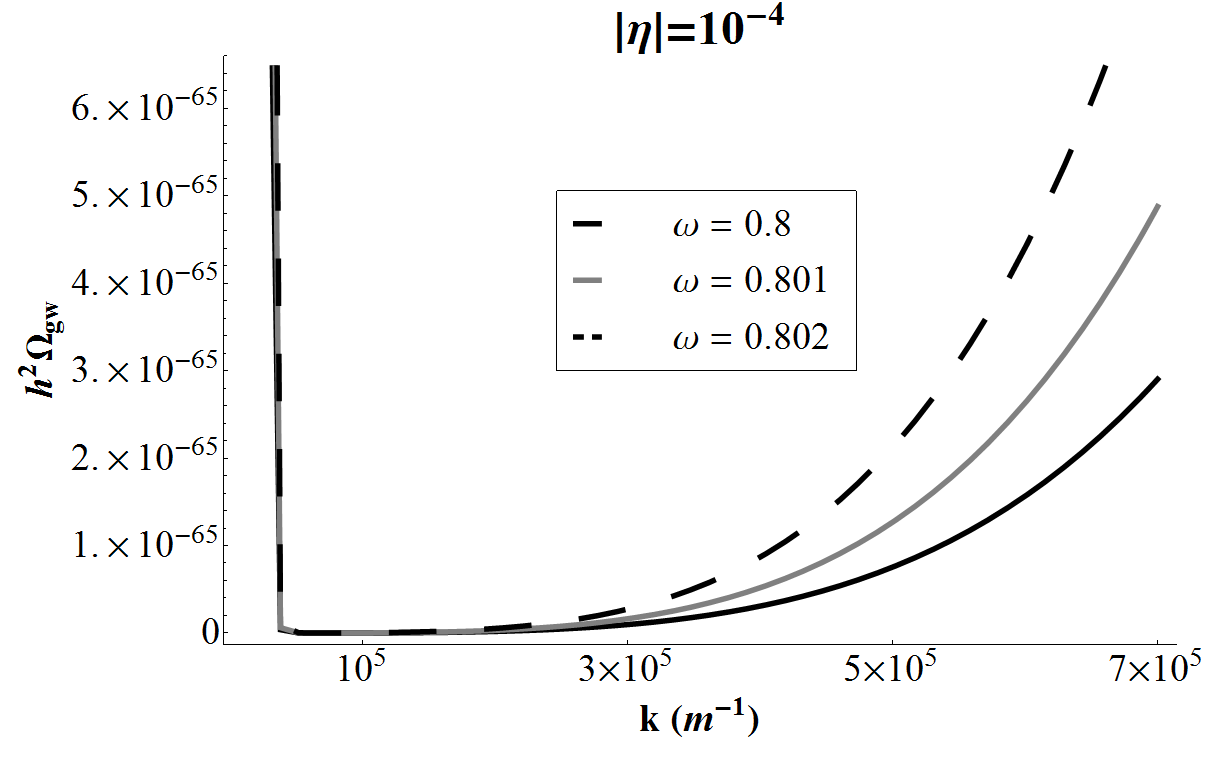}
\end{minipage}
\caption{{The graviton energy density in the Brans-Dicke theory as a function of $k$, for small values of  $\omega$ and fixed $\eta$.}}
\label{fig:densidade1}
\end{figure}
\begin{figure}[!t]
\begin{minipage}[t]{0.48\linewidth}
\includegraphics[width=\linewidth]{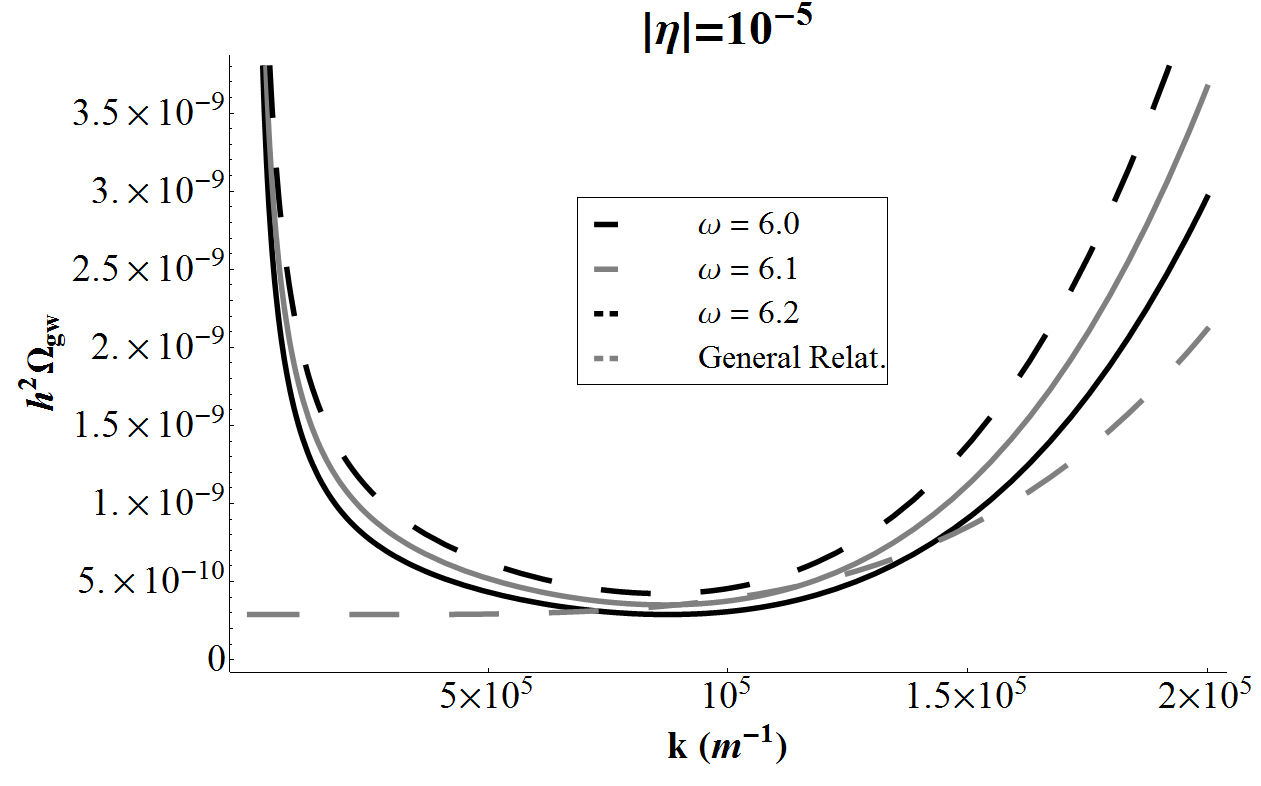}
\end{minipage} \hfill
\begin{minipage}[t]{0.48\linewidth}
\includegraphics[width=\linewidth]{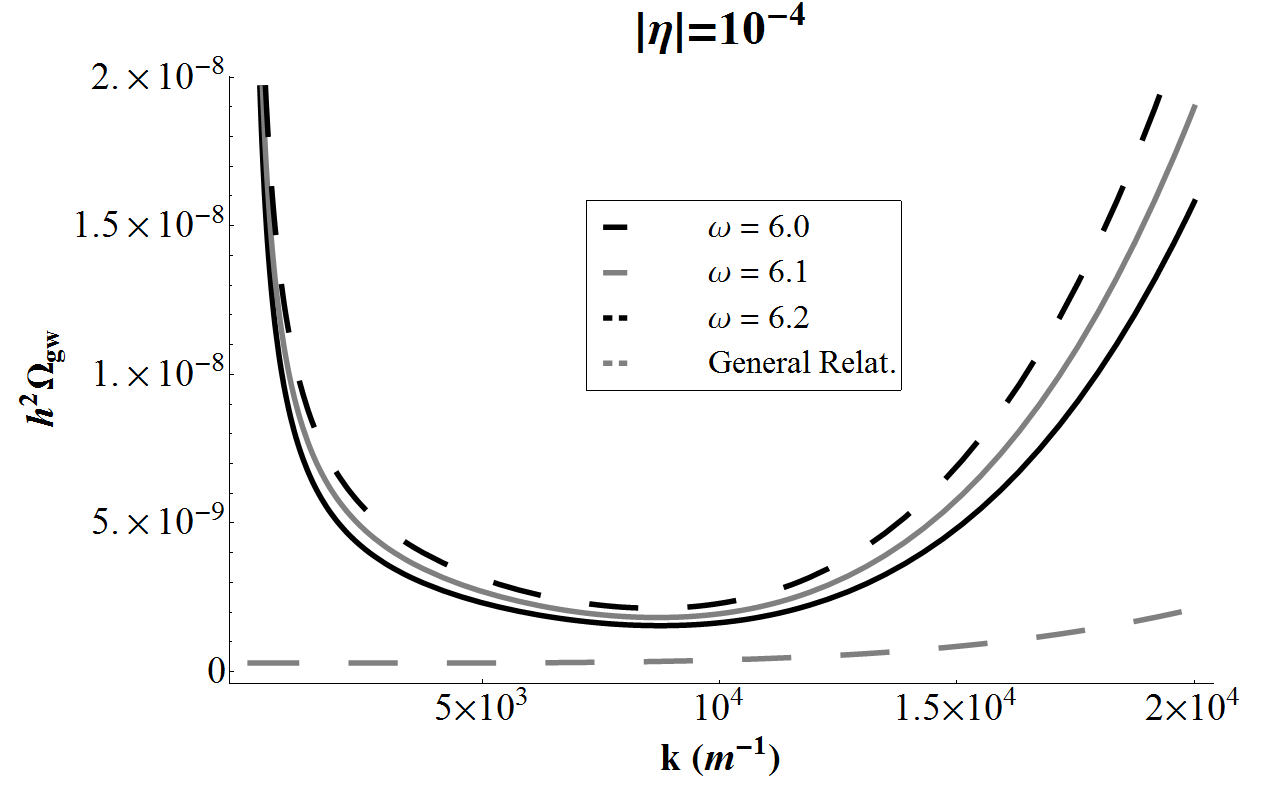}
\end{minipage}
\caption{{The energy density of the gravitational waves in the Brans-Dicke theory as a function of $k$ compared with the energy density in the General Relativity, for some different values of $\omega$ and fixed $\eta$.}}
\label{fig:densidade2}
\end{figure}
\begin{figure}[!t]
\begin{minipage}[t]{0.48\linewidth}
\includegraphics[width=\linewidth]{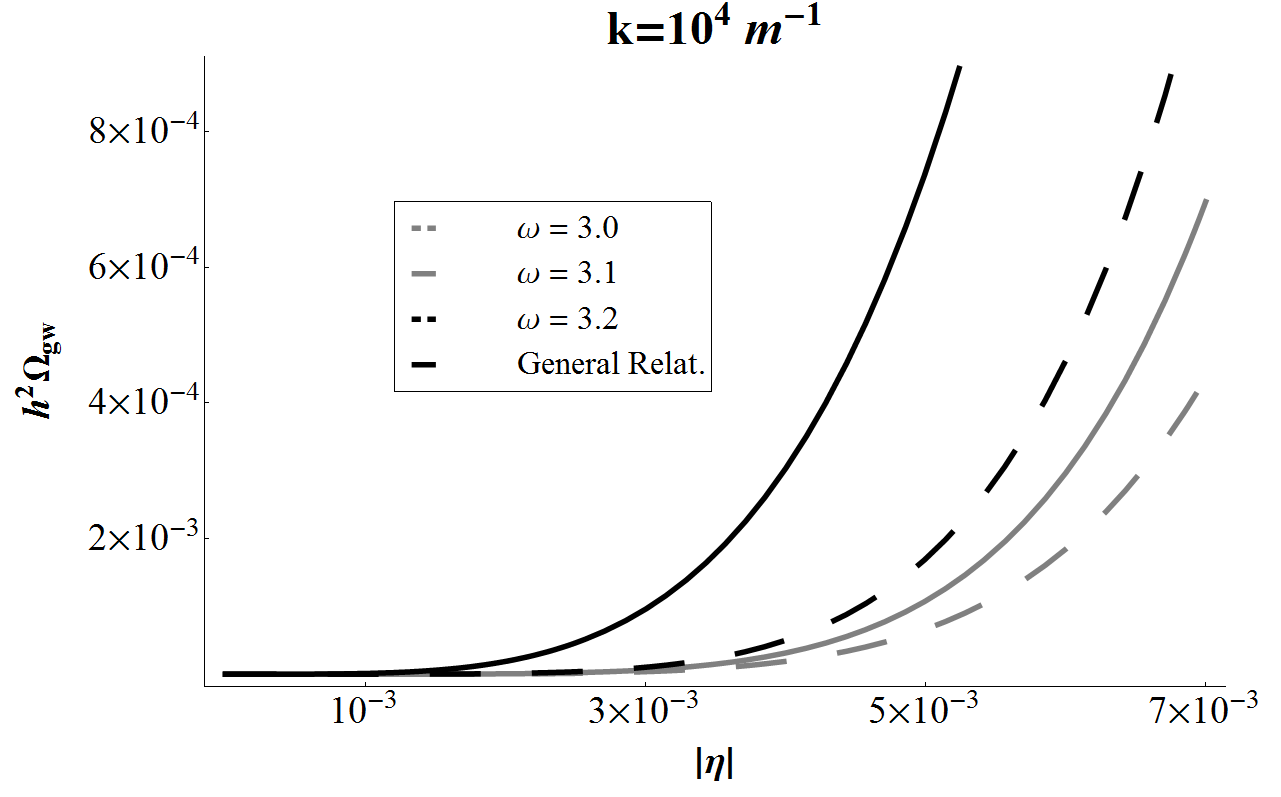}
\end{minipage} \hfill
\begin{minipage}[t]{0.48\linewidth}
\includegraphics[width=\linewidth]{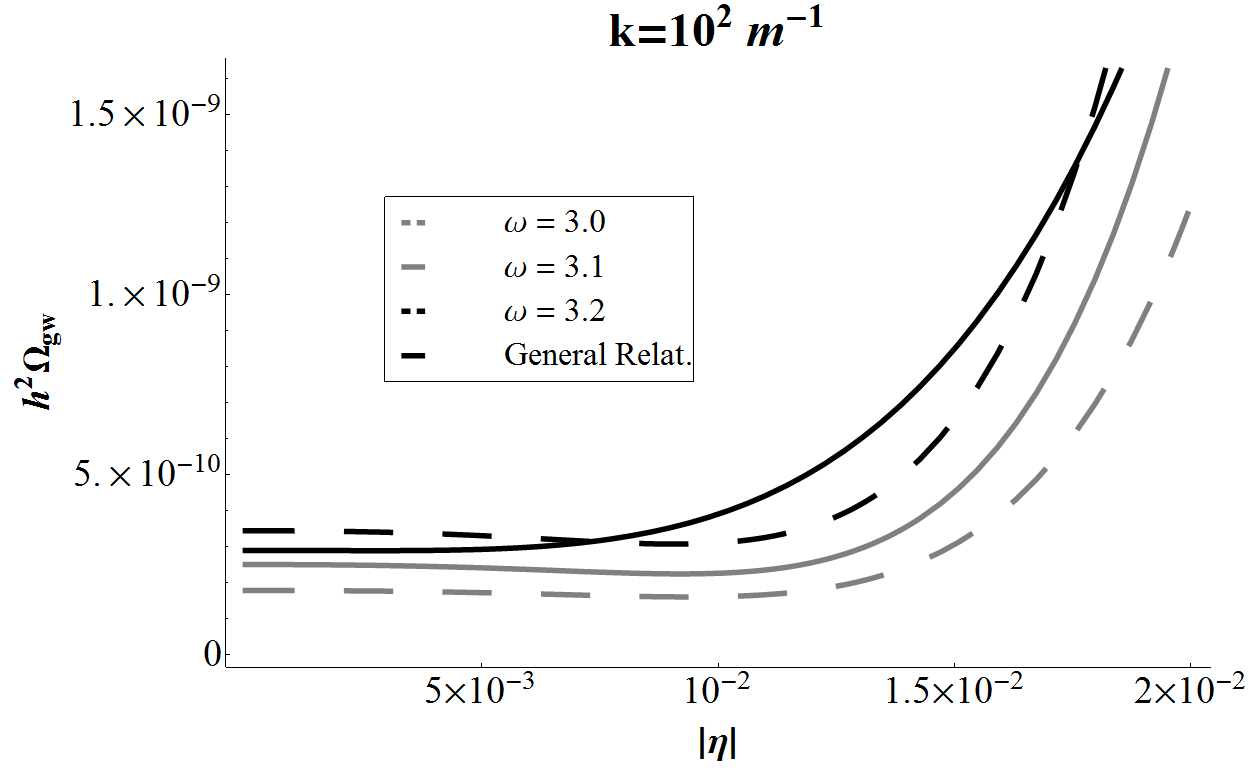}
\end{minipage}
\caption{{The energy density of the gravitational waves in the Brans-Dicke theory as a function of $|\eta|$ compared with the energy density in the General Relativity, for some different values of $\omega$ and with fixed wavenunber $k$.}}
\label{fig:densidade3}
\end{figure}

We stress that when the modes re-enter the horizon the energy density is proportional to the power spectrum. The behavior of the spectral energy density are shown in Fig.~\ref{fig:densidade1} - Fig.~\ref{fig:densidade3}.

\section{Conclusions}
\label{conclusions}
\par
         In this work we have considered the quantum gravitational waves production during inflation in the Brans-Dicke theory. In general, we realize that, in contrast with the General Relativity, not only the scale factor $a(t)$ but also the Brans-Dicke scalar field $\phi(t)$ models the time evolution of the gravitational waves of cosmological origin.
    \par
In the initial inflationary phase of the Universe there is a natural initial vacuum state, from wich the particle occupation number can be determined. We computed the quantum particles number $N_k$ created during extended inflation and the results indicate that the number predicted by this model approximates that predicted by General Relativity when the Brans-Dicke parameter $\omega$ is big enough, as we can see in Fig.~\ref{fig01}~-~Fig.~\ref{fig04}.
    \par
         Afterwards we calculate three observable parameters: the quantum power spectrum $P_k$, the quantum spectral index $n_T$ and the quantum 
         energy density $\Omega(k,\eta)$ of the gravitational waves. The result obtained for the power spectrum $P_k$ shows that, different from the 
         local observations and from the previous result acquired for the graviton number $N_k$, this model is comparable to the General Relativity
         one only if $\omega$ assumes small values like $0.5<\omega<1$. Big values of $\omega$ make the power spectrum to explode. 
    \par
         For the spectral index we obtain a different result. We can compare the outcome of the General Relativity for the spectral index $n_T$ with
         that of the Brans-Dicke theory when $\omega>>1$, result that agree with the one obtained for the quantum particle number $N_k$, and when
         $\omega\rightarrow\infty$ it yields a scale invariant perturbation.
    \par
         In the case of the quatum energy density big values of $\omega$ make $\Omega(k,\eta)$ to explode and the General 
         Relativity prediction is approximated when $1<\omega <10$, depending on the values of the wavenumber and the conformal time. 
    \par
We think the main problem with the scenario of Brans-Dicke is valued at $\omega$. In our study, the values vary significantly from the values obtained by traditional testing sites. The parameter $\omega$ assumes small or large values depending on the observable calculated. However, these drawbacks can be, in principle, circumvented if the gravitational coupling is scale-dependent, as suggested by considering quantum effects \cite{omega01, omega02}.  In this work, we made no attempt to reconcile the values of $\omega$ deducted from local tests, with the corresponding values deduced from the analysis performed here. But, evidently, the problem must be addressed in order to have a completely realistic cosmological scenario. When gravitational waves are detected we can verify if $\omega$ has different values for different cosmological scales.
    \par
A complete study of the Brans-Dicke theory in terms of fluctuations of quantum origin also involves the calculation of scalar perturbations. This is a natural continuation of this work. The results of scalar perturbations during inflation using the Brans-Dicke theory can be compared with the results of General Relativity and with the available observational data, especially data from the cosmic microwave background radiation. We can reduce the number of the variables calculating the ratio between the tensor and the scalar power spectra $r_T =P_T/P_S$ and with this results to obtain direct information about the state of the inflationary universe. An other step to generalize our study is to do the same calculating for both scalar and tensor perturbation but in a model where the Brans-Dicke parameter $\omega$ varies with the scalar field $\phi$, \textit{i. e.}, $\omega=\omega(\phi)$. Such work is in preparation and will be available soon.
\par
\vspace{0.5cm}

{\bf Acknowledgements:} We thank CNPq (Brazil) and CAPES (Brazil) for partial financial support.



\begin{thebibliography}{99}


\bibitem{paul1} R. Crittenden, R.L. Davis and P.J. Steinhardt, Astrophys. J. {\bf 417}, L13 (1993), [arXiv:astro-ph/9306027];
\bibitem{paul2} R. Crittenden, J.R. Bond, R.L. Davis, G. Efstathiou and P.J. Steinhardt,Phys. Rev. Lett. {\bf 71}, 324 (1993), [arXiv:astro-ph/9303014];
\bibitem{mass} M. Giovannini, PMC Phys. {\bf A} 4, 1 (2010), [arXiv:0901.3026v1]; 
\bibitem{gwobs} LIGO: www.ligo.caltech.edu/\quad, \\ 
                VIRGO: www.virgo.infn.it/\quad, \\	
               	MiniGrail: www.minigrail.nl/\quad, \\
	              CLIO: www.icrr.u-tokyo.ac.jp\quad, \\
                GEO 600: www.geo600.org/\quad, \\
	              TAMA 300: tamago.mtk.nao.ac.jp/\quad, \\
	              Gr\'aviton Project: www.das.inpe.br/graviton/index.html\quad, \\	
	              AIGO: www.gravity.uwa.edu.au/\quad, \\
	              LCGT: gw.icrr.u-tokyo.ac.jp/lcgt/\quad, \\	
	              LISA: lisa.nasa.gov/\quad;
\bibitem{uzan} A. Riazuelo and J.P. Uzan, Phys. Rev. {\bf D62}, 083506 (2000), [arXiv:astr-ph/0004156];
\bibitem{ungarelli} A. Buonanno, M. Maggiore and C. Ungarelli, Phys. Rev. {\bf D55}, 3330 (1997), [arXiv:gr-qc/9605072];
\bibitem{sanchez} M.P. Infante and N. S\'anchez, Phys. Rev. {\bf D61}, 083515 (2000), [arXiv:hep-th/9907185];
\bibitem{gasperini} M. Gasperini, Phys. Rev. {\bf D56}, 4815 (1997), [arXiv:gr-qc/9704045];
\bibitem{1} C. Brans and R. H. Dicke, Phys. Rev. {\bf 124}, 925 (1961);
\bibitem{2} P.A.M. Dirac, Proc. Roy. Soc. (London) {\bf A165}, 199 (1938);
\bibitem{3} P. Jordan, Z. Physik {\bf 157}, 112 (1959);
\bibitem{4} G.C. McVittie, M.N.R.A. Soc. {\bf 183}, 749 (1978);
\bibitem{faraoni} V. Faraoni, E. Gunzig and P. Nardone, Fund. Cosmic Phys. {\bf 20}, 12 (1999), [arXiv:gr-qc/9811047v1];
\bibitem{santiago} D.I. Santiago and A.S. Silbergleit, Gen. Rel. Grav. {\bf 32}, 565 (2000), [arXiv:gr-qc/9904003v1];
\bibitem{5} A.H. Guth, Phys. Rev. {\bf D23}, 347 (1981);
\bibitem{cassini} B. Bertotti, L. Iess and P. Tortora, Nature {\bf 425}, 374 (2003);
\bibitem{will} C. M. Will, Living Rev. Relativity {\bf 4}, 4 (2001), [arXiv:gr-qc/0103036];
\bibitem{6} D. La and P. J. Steinhardt, Phys. Rev. Lett. 62, 376 (1989);
\bibitem{7} A. R. Liddle and D. Wands., Phys. Rev. {\bf D} 45, 2665 (1992);
\bibitem{riess} A.G. Riess et al, Astron. J. {\bf 116}, 1009 (1998), [arXiv:astr-ph/9805201];
\bibitem{quinte} R.R. Caldwell, R. Dave and P. J. Steinhardt, Phys. Rev. Lett. {\bf 80}, 1582-1585 (1998), [arXiv:astr-ph/9708069];
\bibitem{k} C. Armendariz-Picon, V. Mukhanov and P. J. Steinhardt, Phys.Rev.Lett. {\bf 85}, 4438-4441 (2000), [arXiv:astr-ph/0004134];
\bibitem{fantasma} V. Faraoni, Phys.Rev. {\bf D69}, 123520 (2004), [arXiv:gr-qc/0404078];
\bibitem{chap} N. Ogawa, Phys.Rev. {\bf D62}, 085023 (2000), [arXiv:hep-th/0003288]; J.C. Fabris, S.V.B. Goncalves and P.E. de Souza, Gen.Rel.Grav. {\bf 34}, 53-63 (2002), [arXiv:astr-ph/0203441]; M. C. Bento, O. Bertolami and A. A. Sen, Phys.Rev. {\bf D66}, 043507 (2002), [arXiv:gr-qc/0202064];
\bibitem{rose} J.C. Fabris, S.V.B. Gon\c calves and R. de Sa Ribeiro, Grav. Cosmol. {\bf 12}, 49-54 (2006), [arXiv:astro-ph/0510779];
\bibitem{barrow} J. D. Barrow, J. P. Mimoso and M. R. G. Maia, Phys. Rev. {\bf D48}, 3630 (1993);
\bibitem{plinio} J.P. Baptista,   J.C. Fabris and S.V.B. Gon\c calves, Astrophysics and Space Science {\bf 246}, 315-331 (1996), [arXiv:gr-qc/9603015];
\bibitem{birrel} N. D. Birrel and P. C. W. Davies, {\it Quantum fields in curved space}, Cambridge University Press, Cambridge (1982);
\bibitem{inflacao} D. Baumann, \textit{TASI Lectures on Inflation}, Lectures da 2009 Theoretical Advanced Study Institute na Univ. of Colorado, Boulder, [arXiv:0907.5424v1];
\bibitem{abramowitz} M. Abramowitz e I. A. Stegun, \textit{Handbook of Mathematical Functions}, National Bureau of Standards, (1964);
\bibitem{dodelson} S. Dodelson, \textit{Modern Cosmology}, Academic Press, (2003);
\bibitem{omega01} M. Reuter and H. Weyer, Phys. Rev. {\bf D70}, 124028 (2004), [arXiv:hep-th/0410117];
\bibitem{omega02} I.L. Shapiro, J. Sola and H. Stefancic, JCAP {\bf 0501}, 012 (2005), [arXiv:hep-ph/0410095].


\end{thebibliography}
\end{document}